\documentclass[%
 reprint,
 amsmath,amssymb,
 aps,
]{revtex4-2}

\usepackage{hyperref}
\usepackage[table]{xcolor}
\usepackage{multirow}
\usepackage{graphicx}
\usepackage{dcolumn}
\usepackage{bm}
\usepackage{siunitx}
\usepackage{lipsum}

\begin{document}

\preprint{APS/123-QED}

\title{Predictors and Socio-Demographic Disparities in STEM Degree Outcomes: \\
A UK Longitudinal Study using Hierarchical Logistic Regression}

\author{Andrew Low and Yasemin Kalender}
\affiliation{%
 Department of Physics, University of Liverpool 
}%

\date{\today}

\begin{abstract} 
Socio-demographic disparities in STEM degree outcomes impact the diversity of the UK’s future workforce, particularly in fields essential for innovation and growth. Despite the importance of institution-level, longitudinal analyses in understanding degree awarding gaps, detailed multivariate and hierarchical analyses remain limited within the UK context. This study addresses this gap by using a multivariate binary logistic model with random intercepts for STEM subjects to analyse predictors of first-class degree outcomes using a nine-year dataset (2013/14 to 2021/22) from a research-intensive Russell Group university. We find that prior academic attainment, ethnicity, gender, socioeconomic status, disability, age, and course duration are significant predictors of achieving a first-class degree, with Average Marginal Effects calculated to provide insight into probability differences across these groups. Key findings reveal that Black students face a significantly lower likelihood of achieving first-class degrees compared to White students, with an average 16\% lower probability, while students graduating from 4-year degree programmes have an average 24\% higher probability of achieving a first-class degree relative to those on 3-year programmes. Although male students received a higher proportion of first-class degrees overall, our multivariate hierarchical model shows higher odds for female students, underscoring the importance of model choice when quantifying awarding gaps. Baseline odds for first-class outcomes rose considerably from 2015/16, peaking in 2020/21, indicating possible grade inflation during the COVID-19 pandemic. Interaction effects between socio-demographic variables and graduation year indicate stability in ethnicity, disability, and socioeconomic awarding gaps but reveal a declining advantage for female students over time. These findings contribute to the literature and provide a transferable, evidence-based framework for departments and institutions to monitor awarding gaps, helping to guide the development of targeted, equity-driven policies that support a diverse and inclusive STEM workforce in the UK.
\end{abstract}

\maketitle


\section{Introduction}

The persistence of socio-demographic disparities in academic outcomes remains a significant barrier to equity in higher education. In the context of STEM degrees, these awarding gaps — differences in academic outcomes based on factors such as ethnicity, gender, socioeconomic status, and age — are particularly concerning. The existence of such gaps not only highlights the presence of structural inequality in higher education, but also limits the diversity and number of qualified professionals in STEM fields, which are critical for innovation and economic growth.

Identifying and addressing these disparities is essential to ensuring that all socio-demographic groups have equal opportunities to succeed in STEM education and related careers. Recent reports by the Industrial Strategy Council and the OECD highlight significant projected skills gaps in STEM fields by 2030 within the UK \cite{IndustrialStrategyCouncil2019, OECD2017, OECD2024}. With over two million jobs projected to rely on physics skills alone \cite{EmsiBurningGlass2022, IoP2020}, the importance of fostering a diverse and skilled STEM workforce cannot be overstated. However, the persistence of STEM awarding gaps exacerbates the skills shortage by restricting the number of students who graduate with first-class or upper second-class degrees and who then choose to pursue a career in STEM.

Early efforts to quantify awarding gaps began in the 1990s and have since developed into a substantial body of research exploring the relationship between factors such as parental education, socioeconomic status, ethnicity, gender, age, and disability \cite{Galindo-Rueda2004, AdvanceHE2023, Boero2022}. This research reflects the evolving demographic profile of undergraduate students. These gaps not only impact the academic success of students from traditionally marginalised backgrounds but also pose additional barriers to recruitment and retention in STEM careers. With the recent establishment of the Department for Science, Innovation, and Technology in the UK and an increased focus on the STEM sector skills gap \cite{NAO2023}, it is crucial for institutions to develop effective strategies to identify and mitigate awarding gaps.

However, despite the high profile nature of the problem, there has been limited institution-level research into first-class STEM degree awarding gaps in the UK. Furthermore, analysis of awarding gaps often suffer from oversimplified statistical models that do not take account of confounding variables or nested data structures \cite{DusenNissen2019, Hubbard2024}. In addition, year-by-year  institutional analysis often suffers from small sample sizes and a lack of statistical power.

This study aims to address these issues by employing hierarchical logistic regression to analyse a large 9-year
dataset from a research-intensive Russell Group university. It seeks to identify predictors of first-class STEM degrees and examine the size and persistence of awarding gaps over time. Ultimately, this study aims to support the UK’s future workforce needs by providing a rigorous analytical framework that other institutions can use to identify awarding gaps, and track the impact of institutional interventions.

\section{Background}

\subsection{Disparities in STEM Outcomes}

The observed gaps in exam performance, pass rates, and graduate outcomes in STEM subjects across different socio-demographic groups are a growing concern for educational institutions. Research into awarding gaps has tended to focus on uncovering the predictors of academic success, including academic and socio-demographic characteristics as well as psychological factors such as student mindset, motivation, and sense of belonging \cite{Li2024, Barnes2021, Cagliesi2023, Griffith2010, Whitcomb2020, Knight2021, Canning2019, Hyater-Adams2019, Kalender2019}. Prior academic achievement in the form of high-school grades is a consistently significant predictor of  academic outcomes at university, even after controlling for other factors \cite{salehi2020, PhysRevPhysEducRes.15.020114, PhysRevPhysEducRes.17.010108, PhysRevPhysEducRes.18.020117, vidalrodeiro2015, SadlerTao2007, Hazari2007}.

Numerous studies have shown that capable students from minority backgrounds face additional disadvantages once on campus, with ethnicity, first-generation status, gender, and socioeconomic status acting as significant predictors of academic performance in STEM courses \cite{Eddy2014b, crisp2009, mervis2010, mervis2011, Greene2008, Cagliesi2023, Griffith2010, AdvanceHE2022, Whitcomb2020, Eddy2014, Madsen2013, Tai2001, Knight2021, Costello2023}. Beyond incoming academic preparation, institutional and environmental factors also impact student experiences and outcomes. For example, student attitudes and engagement \cite{Kalender2020,Burkholder1, PhysRevPhysEducRes.18.020126, adams2006}, local classroom or campus environment \cite{PhysRevPhysEducRes.20.010118, reid2003}, teaching style \cite{hake1998, theobald2020}, assessment structures, and grading practices \cite{PhysRevPhysEducRes.16.020114, PhysRevPhysEducRes.18.020103, PhysRevPhysEducRes.16.020125} have been shown to correlate with academic performance in introductory STEM courses. 

There are also significant socio-demographic disparities in drop-out rates on STEM degree programmes \cite{chen2013}. For example, the education level of a student's parents can have a positive association with student persistence in college, even after controlling for academic preparation \cite{warburton2001}. Evidence suggests that introductory STEM courses may disproportionately drive underrepresented minority students out of these fields \cite{Hatfield2022, Tyson2007, Shaw2010, rieglecrumb2019}. Strategies to improve persistence include orientation programmes, early warning systems, learning communities, and outreach programmes \cite{Sithole2017}. These demographic disparities are not only restricted to undergraduate courses but have also been identified in graduate admissions \cite{PhysRevPhysEducRes.18.020140}.

While most research identifies awarding gaps where traditionally marginalised groups perform worse, exceptions do exist. Some studies have shown that demographic factors such as gender, underrepresented minority status, first-generation college student status, and low socioeconomic status are not significant predictors of academic outcomes \cite{PhysRevPhysEducRes.16.020130, PhysRevPhysEducRes.15.020120, PhysRevPhysEducRes.17.010106, PhysRevPhysEducRes.18.010114}. This highlights the importance of institution level analysis to identify the types, magnitudes, and causes of awarding gaps within a local context.

\subsection{STEM Awarding Gaps in the UK}

An awarding gap in the UK typically refers to the difference in the proportion of students from different socio-demographic groups achieving a first-class or upper second-class degree \cite{Smith2001}. The term `awarding gap' is used rather than `attainment gap' to highlight that the issue is structural and institutional, rather than reflective of a student's capability \cite{AdvanceHE2022}. 

In the UK, university degrees are classified based on the student's overall score at the end of their degree: First-Class Honours (First) requires a score of 70\% or above; Upper Second-Class Honours (2:1) requires a score of 60-69\%; Lower Second-Class Honours (2:2) requires a score of 50-59\%; and Third-Class Honours (Third) requires a score of 40-49\%. Degree classifications, and the perceived quality and prestige of the university attended by the student, are often used by employers as an indicator of a candidate’s academic capabilities, with most graduate jobs requiring at least an upper second-class degree for entry onto graduate programmes. Recent research conducted by the Institute for Fiscal Studies and the Department for Education \cite{Britton2022} indicates that those who achieve a first-class degree earn significantly more five years after graduation, underscoring the importance of understanding the predictors of first-class degree outcomes.

Entry into university in the UK depends on how well a student performs during their final exams at high school. In England, Wales, and Northern Ireland, students typically study 3-5 A-levels, graded from A* (highest) to E (pass), and the grade profile determines the universities a student can apply to and the courses they can study.

The role of incoming academic preparation has been widely studied in the UK context. For example, the number of A* grades achieved in A-Levels is a strong predictor of achieving a first or upper second-class degree, even after accounting for other background characteristics \cite{vidalrodeiro2015}. Advanced mathematics courses, such as A-level Further Mathematics (an optional maths extension A-Level subject), are reported to provide good preparation for engineering degrees \cite{darlington2017engineering}. Studies have also shown differences in academic and labour market outcomes depending on whether the student attended state-funded or privately-funded high schools \cite{Boero2024, Green2017, Green2017}.

Although there have been several multi-institution and multi-subject UK studies exploring the association between high school characteristics and university outcomes \cite{Crawford2014a, HEFCE2003, HEFCE2005, DfE2013}, socio-demographic differences in drop-out rates, degree completion, and degree classification \cite{Crawford2014b, Thiele2015, Andrews2023}, assessment types \cite{Cagliesi2023} and graduate earnings \cite{CrawfordVignoles2014}, there are surprisingly few STEM-specific studies at the institutional level that employ sophisticated statistical modeling to track awarding gaps over time or focus specifically on first-class degree outcomes.

\subsection{Quantitative methods in STEM education}

When attempting to identify the predictors of academic outcomes in STEM subjects, and the existence of awarding gaps, it is essential to take a robust approach to statistical analysis to ensure that any conclusions drawn are valid. Recently, there has been an increased focus on the use and critique of more sophisticated statistical approaches within the STEM education literature, as well as the use of machine learning techniques \cite{Aiken2021, PhysRevPhysEducRes.15.020101, PhysRevPhysEducRes.15.020110, PhysRevPhysEducRes.20.010113, konig2017,baepler2010,baker2009,papamitsiou2014, PhysRevPhysEducRes.17.010119,baepler2010,baker2009}. 

Careful consideration of potentially confounding variables is required when investigating socio-demographic awarding gaps, as the appearance of a gap might disappear once other factors, such as incoming academic preparation, or parental education, are controlled for. Therefore it is essential to use multivariate regression models that account for multiple variables simultaneously  \cite{PhysRevPhysEducRes.15.020114, Burkholder1, Burkholder2}. 

It is also important to take account of hierarchical structure in nested datasets, as students are often grouped into subjects, departments, schools, and universities. Improper use of single-level models can lead to biased findings \cite{DusenNissen2019}. Since degree outcomes in the UK are not measured on a continuous scale, either a multivariate binary logistic or ordinal logistic model is required to account for the categorical nature of the outcome variable whilst allowing for multiple predictor variables \cite{PhysRevPhysEducRes.15.020110}. 

 Despite the wealth of research emphasising the importance of performing multivariate analysis and accounting for hierarchical structure, many institutions in the UK still simply quote percentage differences in the proportion of students who achieve first-class or upper-second-class degrees, asserting that this represents the size of the awarding gap at their institution. In this study, we will directly address this problem by proposing a robust analytical framework that takes into account multiple variables and hierarchical structure, and can be used for identifying and tracking awarding gaps over time.

\section{Research Questions}

This study aims to investigate the academic and socio-demographic predictors of first-class STEM degrees and the presence of STEM awarding gaps at a research-intensive Russell Group university between 2013/14 and 2021/22. The following research questions guide this investigation:

\begin{enumerate}
    \item [\textbf{RQ1:}] What are the pre-university academic and socio-demographic predictors of achieving a first-class STEM degree?  
    \item [\textbf{RQ2:}] Which student demographic groups exhibit STEM degree awarding gaps?
    \item [\textbf{RQ3:}] Have any identified awarding gaps changed significantly in the years following 2013/14?
\end{enumerate}

The first research question aims to identify factors such as pre-university academic achievement, gender, ethnicity, socioeconomic status, and other relevant characteristics that may impact the likelihood of a student achieving a first-class STEM degree at this particular university. The second research question explores whether there is evidence to suggest that certain socio-demographic groups within the dataset are less likely to achieve a first-class degree. The final research question addresses whether the size and significance of any identified awarding gaps have changed in the years following 2013/14.

This study makes several unique contributions to the research literature. First, by focusing on a single research-intensive Russell Group university—part of an association of 24 leading public research universities in the UK known for their commitment to high-quality research and teaching \cite{RussellGroup}—and using a large dataset spanning nine years, it provides valuable insights into the factors associated with achieving a first-class STEM degree. Second, the use of hierarchical logistic regression offers a robust analytical framework for identifying awarding gaps, while accounting for potential confounding variables and nested data structures \cite{DusenNissen2019}. Third, by calculating the average marginal effects of socio-demographic predictors, we present an easily interpretable metric for socio-demographic awarding gaps, facilitating clearer communication of results to policymakers. Finally, by making use of an interaction model and calculating the year-specific average marginal effects of socio-demographic variables, we provide a method for tracking awarding gaps through time. We hope our analytical approach can serve as a template for other institutions to inform their own research on awarding gaps.

\begin{table*}[tp]
\caption{\label{categoric} Descriptions and Level specifications of categorical variables. Category descriptions and levels are based on the HESA categorisation \cite{HESA2024}. Bold font has been used to identify the reference level within each category, which coincides with the most populated level of each category. Percentages are rounded to nearest 0.5\%.}
\begin{ruledtabular}
\begin{tabular}{llll}
\textbf{Variable} & \textbf{Category Description} & \textbf{Levels} & \textbf{\%} \\
\hline 
\noalign{\smallskip} 
\multirow{6}{*}{\textit{\textbf{Ethnicity}}} & \multirow{6}{*}{\parbox{8cm}{\raggedright Represents student-reported ethnicity based on HESA categorisation, and used in research and benchmarking across the HE sector in the UK. We acknowledge the limitations of this crude categorisation.  }} & \textbf{White} & 82.5 \\
 & & Asian & 6.0 \\
 & & Mixed & 3.5 \\
 & & Black & 2.0 \\
 & & Other & 1.0 \\
 & & Unknown & 5.0 \\
\hline
\noalign{\smallskip} 
\multirow{2}{*}{\textit{\textbf{Gender}}} & \multirow{2}{*}{\parbox{8cm}{\raggedright This variable captures the student's biological sex, rather than their self-identified gender. }} & \textbf{Male} & 51.0 \\
 & & Female & 49.0 \\
\hline
\noalign{\smallskip} 
\multirow{2}{*}{\textit{\textbf{Disability}}} & \multirow{2}{*}{\parbox{8cm}{\raggedright Indicates declared disability status; 0 for no declared disability, 1 for declared disability.}} & \textbf{No Known Disability} & 86.5 \\
 & & Disability & 13.5 \\
\hline
\noalign{\smallskip} 
\multirow{3}{*}{\textit{\textbf{SES}}} & \multirow{3}{*}{\parbox{8cm}{\raggedright Socioeconomic status derived from NS-SEC, combining occupational classes 1-4 (Higher SES) and 5-8 (Lower SES).}} & \textbf{Higher SES} & 69.0 \\
 & & Lower SES  & 15.5 \\
 & & Unknown & 15.5 \\
\hline
\noalign{\smallskip} 
\multirow{4}{*}{\textit{\textbf{Parent Ed}}} & \multirow{4}{*}{\parbox{8cm}{\raggedright The data refers to whether students’ parents or guardians have a higher education qualification. If the information was refused, this has been labelled as `Unknown'.}} & \textbf{Yes} & 49.0 \\
 & & No & 40.5 \\
 & & Don't Know & 5.5 \\
 & & Unknown & 5.0 \\
\hline
\noalign{\smallskip} 
\multirow{3}{*}{\textit{\textbf{School}}} & \multirow{3}{*}{\parbox{8cm}{\raggedright Identifies the category of high school the student attended.}} & \textbf{State School} & 85.5 \\
 & & Private School & 9.0 \\
 & & Unknown & 5.5 \\
\hline
\noalign{\smallskip} 
\multirow{2}{*}{\textit{\textbf{Duration}}} & \multirow{2}{*}{\parbox{8cm}{\raggedright Indicates the length of STEM degree course completed.}} & \textbf{3 Years} & 73.5 \\
 & & 4 Years & 26.5 \\
\hline
\noalign{\smallskip} 
\multirow{2}{*}{\textit{\textbf{Age}}} & \multirow{2}{*}{\parbox{8cm}{\raggedright Indicates the age of the qualified student as of 31st July in the year of graduation.}} & \textbf{18-24} & 98.0 \\
 & & 25+ & 2.0 \\
\end{tabular}
\end{ruledtabular}
\end{table*}

\section{Methodology}

\subsection{Data and Terminology}

The data used in this study were obtained from the Joint Information Systems Committee (Jisc), which provided data from the Higher Education Statistics Agency (HESA) and UCAS under license \cite{HESA2024, Jisc2024}. HESA is a UK organisation that works with higher education providers to collect and curate higher education data sources. The dataset consists of socio-demographic information, pre-university academic qualifications, and degree-classification data for students ($N=8965$) from a research-intensive Russell Group university who graduated with a STEM degree between summer 2014 and summer 2022.

The dataset is rich, containing detailed information on \textit{Gender, Ethnicity, Disability, School} (private versus state school), \textit{Parental Ed} (parental education), \textit{SES} (socioeconomic status), \textit{Age} (age upon graduation), \textit{STEM Subject} (the STEM degree subject they studied), \textit{Duration} (number of years in the degree), prior academic qualifications in the form of \textit{Tariff} (UCAS Tariff points), and \textit{Year} (the year in which the student graduated). The primary outcome variable of interest is a binary variable indicating whether a student achieved a first class degree upon graduation, or not. A summary description of the categorical predictor variables used in the research study can be found in Table \ref{categoric} and Table \ref{subjectcategoric}. Within each categorical variable, the most populated group is used as the reference level.

The \textit{Gender} variable refers to the student's biological sex rather than signifying their gender identity. We acknowledge that gender is a complex social construct that extends beyond binary categories, and that this categorisation fails to recognise non-binary and trans identities. Due to the very small sample size, students whose gender was categorised as `other' were not included in the analysis.

The terminology we use to refer to students from different ethnic backgrounds follows the HESA categorisation, including the following categories (ordered by population size): White, Asian, Mixed, Black, Other, and Unknown. We acknowledge that this system may not fully capture the nuanced ethnic identities within our student population; for example, the category ‘Asian’ includes both South Asian and East Asian students, whose outcomes may differ \cite{Hubbard2024}. When referring collectively to students who are not White, we use the term ‘minority ethnic’ to acknowledge the historical marginalisation of these groups \cite{UKDiversityReport2020}, while recognising that this term, too, has its limitations.

Socioeconomic status is classified according to the National Statistics Socio-economic Classification (NS-SEC), which categorises occupations into eight classes based on occupational skill level \cite{HESA2024, Thiele2015}. For this study, we have combined classes 1-4 as ‘Higher SES’ and classes 5-8 as ‘Lower SES’. We acknowledge that this binary categorisation is a simplified measure of socioeconomic status, used here as a compromise to address small sample sizes in individual SES classes.

The \textit{Disability} variable in our dataset is derived from HESA’s categorisation, based on students’ self-reported disabilities. Before the 2022/23 academic year, students could only report a single disability, which may not fully capture the range of conditions a student may experience. Reporting a disability is optional; therefore, this data reflects only those students who chose to disclose their disability status. Students who did not report a disability, as well as those with entries such as ‘Prefer not to say’ or ‘Not available,’ are grouped under ‘No known disability.’ As a result, this dataset may underrepresent the full scope of the disabled student population, as it excludes students who opted not to disclose a disability.

The \textit{School} variable represents the type of high school attended by the student prior to university. In the UK, high schools are broadly divided into two categories: privately funded and state-funded, with approximately 6\% of all school-age children attending private schools \cite{ISC2023}. 

The \textit{Age} variable represents the age of the qualified student as of 31st July in the year they graduated from their STEM degree. We use a binary categorisation for this variable: students who graduated between ages 18 and 24 form the reference category (as they constitute the majority of students), while students aged 25 and over at graduation are classified as mature students.

The \textit{Parental Ed} variable captures whether a student has indicated that any parent or guardian, including natural parents, adoptive parents, stepparents, or guardians, holds a higher education qualification.

The \textit{Duration} variable indicates the length of the degree course completed by the student. In the UK, a standard Bachelor of Science (BSc) degree lasts three years, while a typical integrated master's degree (e.g. MPhys) in a STEM subject lasts four years. The required UCAS Tariff points for entry onto a four-year integrated master's degree programme are typically higher than those for a three-year BSc programme. 

We retained ‘unknown’ categories for \textit{Ethnicity}, \textit{SES}, \textit{School}, and \textit{Parental Ed} to ensure that all available data are considered in the analysis. This approach helps capture potential demographic patterns that might otherwise be overlooked, acknowledging that students with ‘unknown’ values may face unique or unmeasured influences on degree outcomes, which could reflect distinct socio-demographic or institutional factors.

We see from Table \ref{categoric} that minority ethnic students and students with a declared disability comprise a small portion of the student population at this university. Specifically, 6.0\% are Asian, 3.5\% are of Mixed ethnicity, 2.0\% are Black, and 1.0\% are from other ethnic backgrounds, while 13.5\% of students have a declared disability. The gender balance is nearly equal, with 51.0\% male and 49.0\% female. Parental education levels show that 49.0\% of students have parents or guardians with higher education qualifications, while 40.5\% do not. Approximately two-thirds (69.0\%) of students come from households classified as Higher SES, while 15.5\% are from Lower SES backgrounds. While the majority of students at this institution come in with the advantage of family capital, most (85.5\%) attended state schools, compared to 9.0\% who attended private schools. Additionally, 73.5\% of students spent 3 years on their degree programme, while 26.5\% spent 4 years.

\begin{table}[tp]
\caption{\label{subjectcategoric} Distribution of students in the dataset across different STEM degree subjects, showing the percentage of students within each degree subject relative to the total dataset, and the percentage of first-class degrees awarded in each degree subject. Percentages are rounded to nearest 0.5\%.}
\begin{ruledtabular}
\begin{tabular}{lll}
\textbf{STEM Subject} & \textbf{\% Students} & \textbf{\% 1sts} \\
\hline 
\noalign{\smallskip} 
\textbf{Physics} & 8.0 & 50.0 \\
Aero Eng. & 5.5 & 36.0 \\
Bioscience & 10.0 & 39.5 \\
Chemistry & 8.5 & 33.5 \\
Civil Eng. & 4.0 & 50.0 \\
Comp Sci & 6.0 & 49.5 \\
Ecology & 2.0 & 38.5 \\
Elec Eng. & 1.5 & 18.5 \\
Eng. and Mfg. & 1.5 & 43.0 \\
Genetics & 1.0 & 36.5 \\
Maths & 11.0 & 29.5 \\
Mech Eng. & 6.5 & 48.0 \\
Microbio & 2.0 & 29.0 \\
Mol Bio & 2.5 & 35.5 \\
Psych & 26.5 & 21.5 \\
Soft Eng. & 1.5 & 47.0 \\
Zoology & 2.0 & 31.5 \\
\end{tabular}
\end{ruledtabular}
\end{table}

Table \ref{subjectcategoric} provides the distribution of students across different STEM degree subjects, indicating the percentage of students in each subject. Popular subjects such as Psychology, Mathematics, and Bioscience, have higher student counts, while some subject categories such as Genetics, Engineering and Manufacturing, and Software Engineering, have a smaller number of students. We also see that the total proportion of first-class degrees varies by STEM subject, with Physics, Civil Engineering, and Computer Science awarding the highest proportion of first-class degrees between 2013/14 and 2021/22, and Electrical and Electronic Engineering, and Psychology, awarding the lowest proportion.

The only continuous variable in the dataset is \textit{Tariff}, which represents the students pre-university academic achievement in the form of A-Level grades that have been converted into a numerical Tariff score. Table \ref{univariatepredictors} shows the mean Tariff score for each level within every categorical variable. We see that the mean UCAS Tariff score across the entire dataset is $147.6$. For context, a grade profile of A*A*A* would equate to 168 Tariff points, a profile of AAA would equate to 144 UCAS Tariff points, and a grade profile of BBB would equate to 120 UCAS Tariff points. The fact that some students within the dataset scored significantly higher than 168 UCAS Tariff points simply indicates that they scored highly in more than three A-Level subjects. For the purpose of hierarchical logistic modeling, we will mean standardise UCAS Tariff score such that the mean is equal to zero and a one-unit change corresponds to a one standard deviation change.

Finally, the \textit{Year} variable refers to the calendar year in which the student graduated from their STEM degree. Figure \ref{fig:propovertime} illustrates how the proportion of first-class degrees and the mean Tariff score of graduating students have changed over time between 2013/14 and 2021/22 across the entire dataset. While the proportion of first-class degrees has steadily increased, peaking during the COVID-19 pandemic, the mean Tariff score has, in contrast, decreased over the same period.
\subsection{Analytical Approach}

Our analytical approach involves a multi-stage process designed to investigate each research question rigorously, using univariate analysis as an initial exploratory step and hierarchical logistic regression as the primary method.

\textbf{Univariate analysis:} As a preliminary step, we conduct a univariate analysis to explore potential predictors of first-class STEM degree outcomes and examine socio-demographic groups in which awarding gaps may exist. This exploratory analysis includes calculating the proportion of first-class degrees for each level of socio-demographic variables. We apply a z-test of proportions to detect significant differences compared to reference levels and use t-tests to assess differences in mean UCAS Tariff scores across levels of these variables. This initial stage helps guide the development of our hierarchical models by highlighting patterns warranting further investigation.

\textbf{Hierarchical logistic regression:} The primary analysis for RQ1 and RQ2 involves building a series of hierarchical logistic regression models that incorporate multiple predictors to account for potential confounding variables \cite{Osborne2000,Raudenbush1988}. Given that students are grouped by the STEM Degree subject they study, we include a random intercept for the \textit{STEM Subject} variable to capture subject-specific variation \cite{Burstein1980}. We select the best-performing model using the Akaike Information Criterion (AIC) \cite{akaike1974new}, Bayesian Information Criterion (BIC) \cite{schwarz1978estimating}, pseudo-$R^2$ \cite{nakagawa2013r2}, and the Area Under the Receiver Operating Characteristic Curve (AUC) \cite{hanley1982meaning, Howell2010AUC}. To identify statistically significant predictors of first-class degree outcomes, as well as the presence of awarding gaps, we estimate and plot odds ratios for each predictor. We calculate Average Marginal Effects (AMEs) \cite{Leeper2024Vignette, PerraillonForthcoming, Heiss2022, Breen2018, Pampel2021} to show the average difference in predicted probabilities of achieving a first-class degree across socio-demographic groups. While odds ratios help identify the relative likelihood of achieving a first-class degree across groups, AMEs provide an interpretable metric that translates these differences into absolute changes in probability, allowing us to more intuitively quantify the awarding gaps across socio-demographic groups. This approach strengthens our analysis by offering a clearer, population-wide perspective on the practical impact of each predictor, making it easier to communicate findings to both academic and policy-focused audiences.

\textbf{Inclusion of Interaction Terms:} To address RQ3, we extend our best hierarchical model to include interaction terms. For each socio-demographic variable (e.g. \textit{Gender}, \textit{Ethnicity}, \textit{SES}), we introduce an interaction with \textit{Year} in a series of separate models, treating 2013/14 as the reference year. For instance, to assess shifts in the gender awarding gap, we add a \textit{Gender*Year} interaction term, enabling us to track annual changes in the odds of female students achieving a first-class degree relative to male students. This process is repeated for each socio-demographic predictor in its own model. From each interaction model, we calculate year-specific Average Marginal Effects to capture annual differences in the average predicted probability of achieving a first-class degree across socio-demographic groups. Visualising these AMEs over time provides a clear view of trends in awarding gaps.

\subsection{Statistical Models}

\subsubsection{\label{Univariate}Univariate Analysis}

The proportion of students who achieve a first-class degree within each level of each categorical predictor variable is used to calculate the odds of achieving a first-class STEM degree. The odds ratio (OR) is defined as:
\begin{equation}
OR = \frac{\text{Odds (Achieve 1st} \mid \text{Level)}}{\text{Odds (Achieve 1st} \mid \text{Reference Level)}}
\label{odds}
\end{equation}
where for a single-predictor model we have:
\begin{equation}
\text{Odds} = \frac{p}{1 - p}
\end{equation}
where \(p\) is the proportion of students achieving a first-class degree. To estimate these proportions for each level of each categorical predictor, we construct a binary logistic regression model. Assuming \(i\) indexes students and \(n\) indexes the levels of the categorical predictor variable (excluding the reference level), the single-predictor logistic model is given by:
\begin{equation}
\log\left(\frac{p_i}{1-p_i}\right) = \beta_0 + \sum_{n} \beta_n X_{in}
\label{singlepredictor}
\end{equation}
where \( \beta_0 \) is the intercept and represents the log-odds of achieving a first-class STEM degree for the reference level of the categorical variable. The terms $ \beta_1, \beta_2, \ldots, \beta_n $ are the coefficients associated with each of the $ n $ levels of the categorical predictor variable, with $ X_{in} $ being a binary dummy variable that equals 1 if student $ i $ belongs to level $ n $ of the predictor variable and 0 otherwise. The odds ratios are calculated by exponentiating the beta coefficients, $ \exp(\beta_n) $, and measure the multiplicative change in the odds of achieving a first-class degree when a student belongs to level $n$ of the categorical variable as compared to when they belong to the reference level.

To test the statistical significance of the differences in proportions between each level and the reference level, a z-test of proportions is used. This test assesses whether the observed differences in the proportion of first-class degrees are greater than what might be expected by chance alone \cite{Howell2010AUC}, based on the standard error of the proportion. These univariate analyses provide insights into the individual effect of each demographic variable on the likelihood of obtaining a first-class degree before considering more complex, multivariate relationships in subsequent models.

As part of the univariate analysis, we also explore the distribution of pre-university Tariff scores across different levels within each categorical variable. Welch's t-test is used to compare the average Tariff score across different levels within each category to the reference level. This approach relaxes the assumption of equal variances, offering a more robust framework for statistical analysis.

\subsubsection{Hierarchical Logistic Regression Models}

To address RQ1 and RQ2, we build upon the univariate analysis and construct a hierarchical logistic regression model. This model identifies the academic and socio-demographic predictors of first-class STEM degree outcomes and the existence of awarding gaps. The dataset consists of students (\( i \)) grouped within STEM subjects (\( j \)). The hierarchical structure is modeled with a random intercept for the \textit{STEM Subject} variable to account for subject-level variation. The model specification is as follows:

\begin{equation}
    \log\left(\frac{p_{ij}}{1 - p_{ij}}\right) = (\gamma_{00} + u_{0j}) + \beta_z Z_{ij} + \sum_{k=1}^K \sum_{n=1}^N \beta_{kn} X_{kijn}
\end{equation}
where:
\begin{itemize}
    \item \( p_{ij} \) is the probability of student \( i \) in subject \( j \) achieving a first-class degree.
    \item \( \gamma_{00} \) is the grand mean intercept, representing the average log-odds of achieving a first-class degree when all predictors are at their reference levels.
    \item \( u_{0j} \) is the random intercept for subject \( j \), capturing subject-specific deviations from the grand mean \( \gamma_{00} \).
    \item \( Z_{ij} \) represents the standardised UCAS Tariff score for student \( i \) in subject \( j \).
    \item \( X_{kijn} \) is a binary indicator variable representing whether student \( i \) in subject \( j \) falls into level \( n \) of categorical predictor \( k \) (e.g., gender, ethnicity).
    \item \( \beta_{kn} \) is the coefficient for level \( n \) of categorical predictor \( k \).
\end{itemize}

This hierarchical model structure allows for the examination of both the fixed effects of individual-level predictors and the variability of outcomes introduced by subject-level differences. By including random intercepts $ u_{0j} $, we capture the variation in the log odds of achieving a first-class honours degree that is attributable to the baseline differences between STEM degree subjects within the dataset. We also calculate the intra-class correlation coefficient (ICC) which measures the proportion of the total variance ($\tau^2+\sigma^2$) attributed to the between-subject variance ($\tau^2$):

\begin{equation}
ICC=\frac{\tau^2}{\tau^2+\sigma^2}
\label{icceqn}
\end{equation}

The ICC ranges from 0 to 1, with higher values indicating greater variability between groups \cite{McGrawWong1996}.
Although we explored the inclusion of random slopes within our model during our preliminary analysis, we found that the added model complexity is not merited by a significant increase in model fit or predictive power, and so we do not include random slopes in our model specification. 

\subsubsection{Interaction Models}

To address RQ3, we extend our best hierarchical model by systematically adding interaction terms between the \textit{Year} variable and each socio-demographic predictor to assess whether the effect of these predictors has changed over time. The introduction of interaction terms allows us to examine the evolution of awarding gaps over time. The model specification with interaction terms is as follows:

\begin{align}
    &\log\left(\frac{p_{ij}}{1 - p_{ij}}\right) = (\gamma_{00} + u_{0j}) + \beta_z Z_{ij} \nonumber \\
    &+ \sum_{k=1}^K \sum_{n=1}^N \left( \beta_{kn} X_{kijn} + \beta_{kn, \text{Year}} (X_{kijn} \times \text{Year}_i) \right)
\end{align}

where \( \beta_{kn, \text{Year}} \) is the coefficient for the interaction term between \textit{Year} and level \( n \) of socio-demographic predictor \( k \) and \( X_{kijn} \times \text{Year}_i \) is the interaction between the categorical variable \( X_{kijn} \) (whether student \( i \) in subject \( j \) belongs to level \( n \) of predictor \( k \)) and the \textit{Year} variable. The interaction term captures the time-varying effect of each socio-demographic variable. Statistically significant interaction terms indicate that the relationship between a socio-demographic variable and the likelihood of achieving a first-class degree has changed relative to 2013/14 (the reference year).

To implement our hierarchical model analysis, we use the \texttt{glmer} function from the \texttt{lme4} package in R \cite{bates2015}, which models the log-odds of the outcome as a function of both individual-level and group-level predictors.

\subsubsection{Calculation of Average Marginal Effects (AMEs)}

Average Marginal Effects for continuous variables are calculated by averaging the marginal effects of a predictor variable across all observations in the dataset \cite{Leeper2024Vignette, PerraillonForthcoming, Heiss2022, Breen2018, Pampel2021}. The marginal effect for an individual observation represents the change in the predicted probability of the outcome associated with a small change in the predictor, holding all other variables constant. For a continuous predictor variable \(X_k\), such as \textit{Tariff}, the AME is calculated as:

\begin{equation}
AME(X_k) = \frac{1}{N} \sum_{i=1}^{N} \frac{\partial P(Y_i = 1 \mid X_{ki})}{\partial X_{ki}}
\label{ame_eqn}
\end{equation}

where \(N\) is the total number of observations, and \(\frac{\partial P(Y_i = 1 \mid X_{ki})}{\partial X_{ki}}\) is the partial derivative of the predicted probability with respect to the predictor \(X_{ki}\) for observation \(i\). The AME represents the average effect of a small change in the continuous predictor variable on the probability of the outcome across the entire dataset.

For categorical variables, the AME for a specific level of a categorical variable represents the average difference in the predicted probability of the outcome when the variable is set to that level versus the reference level, across all individuals in the dataset. Take, for instance, the categorical predictor \textit{Ethnicity} with levels Black and White (where White serves as the reference category). The AME for Black students is calculated as the average difference in the predicted probability of the outcome if each individual’s ethnicity were hypothetically set to Black, compared to if it were set to White, regardless of their actual ethnicity. Mathematically, this is expressed as:

\begin{equation}
\begin{aligned}
AME(\text{Black}) &= \frac{1}{N} \sum_{i=1}^{N} \left[ P(Y_i = 1 \mid \text{Black}) \right.\\
&\phantom{=} \left. - P(Y_i = 1 \mid \text{White}) \right]
\end{aligned}
\label{ame_ethnicity_eqn}
\end{equation}
where $P(Y_i = 1 \mid \text{Black})$ is the predicted probability that individual $i$ achieves a first-class degree when their ethnicity is set to Black, and $P(Y_i = 1 \mid \text{White})$ is the predicted probability that the same individual achieves a first-class degree when their ethnicity is set to White. This approach isolates the effect of Ethnicity by evaluating the variable’s impact within the context of the actual dataset.

To investigate how average marginal effects for socio-demographic variables evolve over time, we calculate year-specific AMEs using the output from each interaction model. This process involves setting the year variable to a fixed value for all individuals in the dataset, allowing us to calculate the AME as if the entire dataset were observed in that particular year. The year is then changed and the process repeated. For example, to track changes in the impact of ethnicity on the probability of achieving a first-class degree, we calculate a separate AME for ethnicity for each year in the dataset. This approach enables us to explore how the awarding gap between Black and White students—represented by the AME—shifts over time.


To implement the AME analysis, we use the \texttt{margins} package in R \cite{leeper2024margins}, which provides a straightforward way to compute AMEs for logistic regression models. We then plot the results to provide a helpful visualisation.

\begin{table*}[]
\begin{ruledtabular}
\caption{\label{univariatepredictors} Univariate analysis showing the proportion of first-class degrees awarded at each level of categorical variable, with the reference level highlighted in bold. Included are the z-scores for comparing proportions between each individual level and the reference level, along with the odds ratios. The table also provides the mean tariff scores and results from Welch's two-sample t-test to test the hypothesis of equal means between each level and the reference level. Significance levels are represented by * (\(p<0.05\)), ** (\(p<0.01\)), and *** (\(p<0.001\)).}
\label{univariate}
\begin{tabular}{llllllll}
\textbf{Variable}    & \textbf{Levels}              & \textbf{1st ($\%$)} & \textbf{Z-score} & \textbf{Odds Ratio} & \textbf{\begin{tabular}[c]{@{}l@{}}Mean \\ Tariff\end{tabular}} & \textbf{SD} & \textbf{t-value} \\  
\hline 
\noalign{\smallskip} 
\textit{\textbf{Ethnicity}}   & \textbf{White}               & 36.0                                                                         &                  &                     & 149.3                                                           & 31.9        &                  \\
\textbf{}            & Asian                        & 27.4                                                                         & -4.00***         & 0.67 (0.55-0.81)    & 140.1                                                           & 32.2        & -6.36***         \\
\textbf{}            & Black                        & 19.7                                                                         & -4.33***         & 0.43 (0.29-0.63)    & 133.5                                                           & 35.7        & -5.79***         \\
\textbf{}            & Mixed                        & 30.6                                                                         & -2.01*           & 0.78 (0.61-0.99)    & 142.8                                                           & 31.1        & -3.70***         \\
\textbf{}            & Other                        & 28.8                                                                         & -1.35            & 0.72 (0.43-1.15)    & 141.5                                                           & 32.6        & -2.14*           \\
\textbf{}            & Unknown                      & 29.8                                                                         & -2.95**          & 0.72 (0.58-0.90)    & 137.3                                                           & 49.6        & -4.97***         \\ \\
\textit{\textbf{Gender}}      & \textbf{Male}                & 36.4                                                                         &                  &                     & 144.8                                                           & 34.7        &                  \\
\textbf{}            & Female                       & 32.7                                                                         & -3.77***         & 0.85 (0.78-0.92)    & 150.5                                                           & 31.5        & 8.13***          \\ \\
\textit{\textbf{Disability}}  & \textbf{No Disability}        & 35.2                                                                         &                  &                     & 148.3                                                           & 33.1        &                  \\
\textbf{}            & Disability                   & 30.7                                                                         & -2.97**          & 0.82 (0.72-0.93)    & 142.6                                                           & 33.9        & -5.43***         \\ \\
\textit{\textbf{SES}}         & \textbf{Higher SES}           & 36.0                                                                         &                  &                     & 148.5                                                           & 31.1        &                  \\
\textbf{}            & Lower SES                    & 31.0                                                                         & -3.52***         & 0.80 (0.71-0.91)    & 145.7                                                           & 34.2        & -2.83**          \\
\textbf{}            & Unknown                      & 31.9                                                                         & -2.84**          & 0.83 (0.74-0.95)    & 145.3                                                           & 41.0        & -2.72**          \\ \\
\textit{\textbf{Parent Ed}}   & \textbf{Yes}                 & 36.2                                                                         &                  &                     & 148.8                                                           & 33.0        &                  \\
\textbf{}            & No                           & 33.5                                                                         & -2.57*           & 0.89 (0.81-0.97)    & 147.2                                                           & 33.2        & -2.28*           \\
\textbf{}            & Don't Know                   & 31.7                                                                         & -1.98*           & 0.82 (0.66-0.99)    & 142.8                                                           & 35.2        & -3.57***         \\
\textbf{}            & Unknown                      & 30.8                                                                         & -2.31*           & 0.79 (0.64-0.96)    & 143.7                                                           & 33.5        & -3.15**          \\ \\
\textit{\textbf{School}}      & \textbf{State School}        & 35.1                                                                         &                  &                     & 148.9                                                           & 32.4        &                  \\
\textbf{}            & Private School               & 31.2                                                                         & -2.23*           & 0.84 (0.72-0.98)    & 143.4                                                           & 26.6        & -5.40***         \\
\textbf{}            & Unknown                      & 32.3                                                                         & -1.26            & 0.88 (0.73-1.07)    & 134.6                                                           & 49.5        & -6.34***         \\ \\
\textit{\textbf{Duration}}    & \textbf{3 Years}             & 26.5                                                                         &                  &                     & 144.9                                                           & 32.3        &                  \\
\textbf{}            & 4 Years                      & 57.2                                                                         & 26.09***         & 3.69 (3.34-4.07)    & 155.2                                                           & 34.7        & 12.65***         \\ \\
\textit{\textbf{Age}}         & \textbf{18-24}               & 34.6                                                                         &                  &                     & 148.3                                                           & 32.5        &                  \\
\textbf{}            & 25 plus                      & 36.7                                                                         & 0.58             & 1.10 (0.80-1.50)    & 111.8                                                           & 50.7        & -9.31***         \\ \\
\textit{\textbf{Year}}        & \textbf{2013/14}             & 27.4                                                                         &                  &                     & 161.1                                                           & 35.8        &                  \\
\textbf{}            & 2014/15                      & 30.2                                                                         & 1.46             & 1.18 (0.95-1.46)    & 158.3                                                           & 35.1        & -1.58            \\
\textbf{}            & 2015/16                      & 35.5                                                                         & 3.67***          & 1.49 (1.21-1.85)    & 157.8                                                           & 35.2        & -1.80            \\
\textbf{}            & 2016/17                      & 37.2                                                                         & 4.57***          & 1.61 (1.31-1.98)    & 153.2                                                           & 31.5        & -4.84***         \\
\textbf{}            & 2017/18                      & 33.9                                                                         & 3.21**           & 1.39 (1.14-1.71)    & 146.1                                                           & 32.1        & -9.26***         \\
\textbf{}            & 2018/19                      & 33.6                                                                         & 3.10**           & 1.37 (1.13-1.68)    & 142.5                                                           & 30.9        & -11.73***        \\
\textbf{}            & 2019/20                      & 33.8                                                                         & 3.18**           & 1.38 (1.13-1.69)    & 142.2                                                           & 30.4        & -12.03***        \\
\textbf{}            & 2020/21                      & 41.5                                                                         & 6.45***          & 1.92 (1.58-2.35)    & 141.1                                                           & 31.9        & -12.30***        \\
\textbf{}            & 2021/22                      & 35.9                                                                         & 4.04***          & 1.52 (1.24-1.87)    & 136.1                                                           & 29.4        & -15.82***        \\  
\noalign{\smallskip} 
\hline
\noalign{\smallskip} 
\textbf{All Data}            &  &   \textbf{34.6} & &  &  \textbf{147.6} & \textbf{33.3} & \\
\end{tabular}
\end{ruledtabular}
\end{table*}

\section{Results}

\subsection{Preliminary Patterns in Degree Outcomes}

We begin by presenting the results of our univariate analysis, which examines differences in the proportion of students obtaining a first-class degree across various levels of categoric variables. Using z-tests of proportions and calculating odds ratios, we identified statistically significant differences across multiple groups, as summarised in Table \ref{univariate}.

\textit{\textbf{Ethnicity:}} The results show a statistically significant difference in the proportion of students achieving a first-class degree across different ethnic groups compared to White students. Notably, Black students have 0.43 (95\% CI: 0.29-0.63) times the odds of achieving a first-class degree compared to White students, and Asian students have 0.67 (95\% CI: 0.55-0.81) times the odds. We also note that White students are observed to enter university with statistically significantly higher mean Tariff scores compared to all other ethnic groups. This pattern may suggest a correlation between entry qualifications and degree outcomes, which merits further investigation when building multivariate models.

\textit{\textbf{Gender:}} We observe that the proportion of female students achieving a first-class degree is statistically significantly less than that of male students, with an odds ratio of 0.85 (95\% CI: 0.78-0.92). Interestingly, this occurs despite female students entering university with statistically significantly higher mean Tariff scores compared to male students (150.5 versus 144.8). 

\textit{\textbf{Disability:}} Students with a declared disability are statistically less likely to achieve a first-class degree, with an odds ratio of 0.82 (95\% CI: 0.72–0.93) compared to students without a known disability. This disparity may be partly attributed to the significantly lower mean Tariff scores for students with a declared disability (142.6 vs. 148.3), which will be examined further in the multivariate analysis.

\begin{figure}[tp]
        \centering
        \includegraphics[width=\columnwidth]{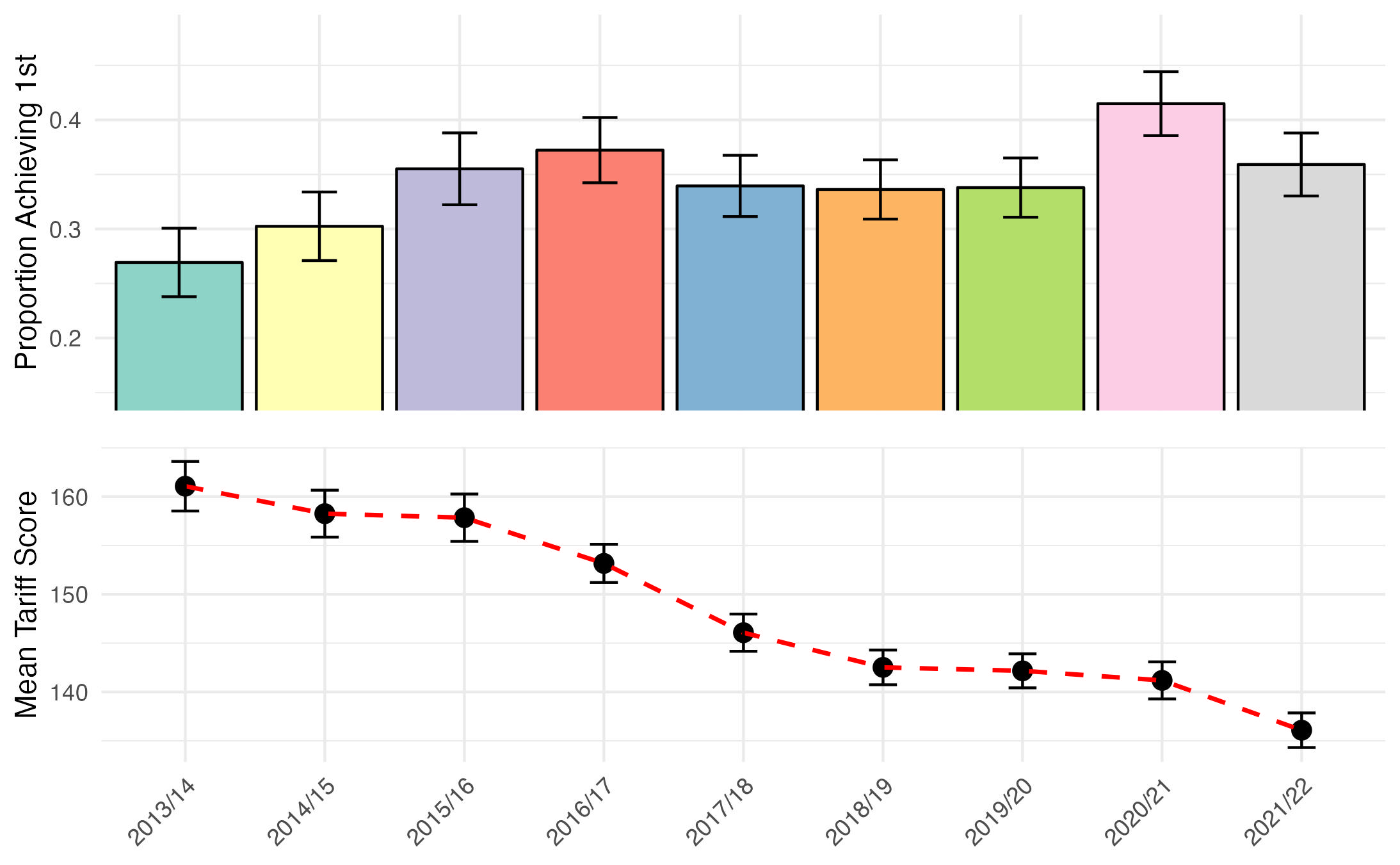}
        \caption{A plot to show how the proportion of first-class degrees, and the mean Tariff score of a graduating student, has changed over time. The error bars in the top plot refer to the 95\% confidence intervals calculated using the standard error on the proportion. The error bars on the bottom plot represent 95\% confidence intervals calculated using the standard error on the mean.}
        \label{fig:propovertime}
 
\end{figure}

\textit{\textbf{Socio-economic factors:}} Students from lower SES families and students whose parents did not attend university achieve statistically significantly lower proportions of first-class degrees, with odds ratios of 0.80 (95\% CI: 0.71-0.91) and 0.89 (95\% CI: 0.81-0.97), respectively. Again, it is worth noting that students from lower SES families have statistically significantly lower mean Tariff scores than those from higher SES families (145.7 compared to 148.5). Additionally, students from private schools show a lower proportion of first-class degrees with an odds ratio of 0.84 (95\% CI: 0.72-0.98), and lower mean Tariff scores compared to those from state schools, challenging common perceptions about educational advantage from private schooling.

\begin{table*}[tp]
\centering
\caption{Summary of Fixed Effects, Odds Ratios, and Average Marginal Effects for best hierarchical logistic model. Fixed effect estimates are provided with standard errors in parentheses, followed by odds ratios with 95\% confidence intervals. The table also provides AME estimates with standard errors and 95\% confidence intervals in parentheses. Significance levels are represented by * (\(p<0.05\)), ** (\(p<0.01\)), and *** (\(p<0.001\)).}
\label{tab:combined_model_summary}
\begin{ruledtabular}
\begin{tabular}{lllll}
\textbf{Effect} & \textbf{Fixed Effects Estimate (SE)} & \textbf{Odds Ratio (95\% CI)} & \textbf{AME Estimate (SE)} & \textbf{AME 95\% CI} \\
\hline
\noalign{\smallskip} 
Ethnicity: Asian & -0.50 (0.11)*** & 0.61 (0.49-0.75) & -0.09 (0.02)*** & (-0.13, -0.06) \\
Ethnicity: Black & -0.94 (0.21)*** & 0.39 (0.26-0.59) & -0.16 (0.03)*** & (-0.22, -0.10) \\
Ethnicity: Mixed & -0.27 (0.13)* & 0.77 (0.59-0.99) & -0.05 (0.02)* & (-0.10, -0.00) \\
Ethnicity: Other & -0.41 (0.27) & 0.66 (0.39-1.12) & -0.08 (0.05) & (-0.17, 0.01) \\
Ethnicity: Unknown & -0.31 (0.14)* & 0.74 (0.57-0.96) & -0.06 (0.02)* & (-0.11, -0.01) \\
Disability: Yes & -0.19 (0.07)** & 0.83 (0.71-0.95) & -0.04 (0.01)** & (-0.06, -0.01) \\
SES: Lower & -0.17 (0.07)* & 0.84 (0.74-0.96) & -0.03 (0.01)* & (-0.06, -0.01) \\
SES: Unknown & -0.09 (0.08) & 0.91 (0.78-1.06) & -0.02 (0.02) & (-0.05, 0.01) \\
Gender: Female & 0.34 (0.06)*** & 1.40 (1.25-1.58) & 0.07 (0.01)*** & (0.04, 0.09) \\
Duration: 4 Years & 1.11 (0.06)*** & 3.04 (2.71-3.42) & 0.24 (0.01)*** & (0.21, 0.27) \\
Tariff & 0.39 (0.03)*** & 1.48 (1.41-1.56) & 0.08 (0.01)*** & (0.07, 0.09) \\
Age: 25+ & 0.51 (0.18)** & 1.66 (1.17-2.36) & 0.10 (0.04)** & (0.03, 0.18) \\
Year: 2014/15 & 0.10 (0.12) & 1.10 (0.87-1.39) & 0.02 (0.02) & (-0.03, 0.07) \\
Year: 2015/16 & 0.46 (0.12)*** & 1.58 (1.25-1.99) & 0.08 (0.02)** & (0.04, 0.12) \\
Year: 2016/17 & 0.67 (0.11)*** & 1.96 (1.57-2.45) & 0.12 (0.02)*** & (0.08, 0.16) \\
Year: 2017/18 & 0.68 (0.11)*** & 1.98 (1.59-2.47) & 0.12 (0.02)*** & (0.08, 0.16) \\
Year: 2018/19 & 0.63 (0.11)*** & 1.87 (1.50-2.33) & 0.11 (0.02)*** & (0.07, 0.15) \\
Year: 2019/20 & 0.61 (0.11)*** & 1.84 (1.48-2.29) & 0.11 (0.02)** & (0.07, 0.15) \\
Year: 2020/21 & 0.94 (0.11)*** & 2.57 (2.06-3.21) & 0.18 (0.02)*** & (0.14, 0.22) \\
Year: 2021/22 & 0.76 (0.11)*** & 2.14 (1.70-2.67) & 0.14 (0.02)*** & (0.10, 0.18) \\ 
\hline
\noalign{\smallskip} 
\multicolumn{2}{l}{\textbf{Model Fit Statistics and Random Effects:}} \\
\noalign{\smallskip} 
\textbf{SD} = 0.41  & \textbf{Pseudo $R^2$ (fixed)} = 0.13 & \textbf{AIC} = 10469.00 & \textbf{AUC} = 0.72 \\
\textbf{ICC} = 0.05 & \textbf{Pseudo $R^2$ (total)} = 0.18 & \textbf{BIC} = 10632.70 & \textbf{logLik} = -5140.1 \\
\end{tabular}
\end{ruledtabular}
\end{table*}

\textit{\textbf{Duration of degree programme:}} There is a statistically significant difference in the likelihood of students achieving a first-class degree between 4-year and 3-year programmes. Students from 4-year programmes have 3.69 times the odds (95\% CI: 3.34-4.07) compared to their peers from 3-year programmes. This may, in part, be explained by the statistically significantly higher mean UCAS Tariff scores of students on four-year courses compared to three-year courses (155.2 versus 144.9). 

\textit{\textbf{Year of graduation:}} There is a statistically significant increase in the proportion of students achieving first-class degrees in each year from 2015/16 onward, relative to 2013/14. The odds ratio for achieving a first-class degree peaked in 2020/21, following the COVID-19 pandemic, at 1.92 (95\% CI: 1.58–2.35), suggesting potential grade inflation during this period.

\begin{figure*}[tp]
    \centering
    \begin{minipage}{0.48\textwidth}
        \centering
        \includegraphics[width=\linewidth]{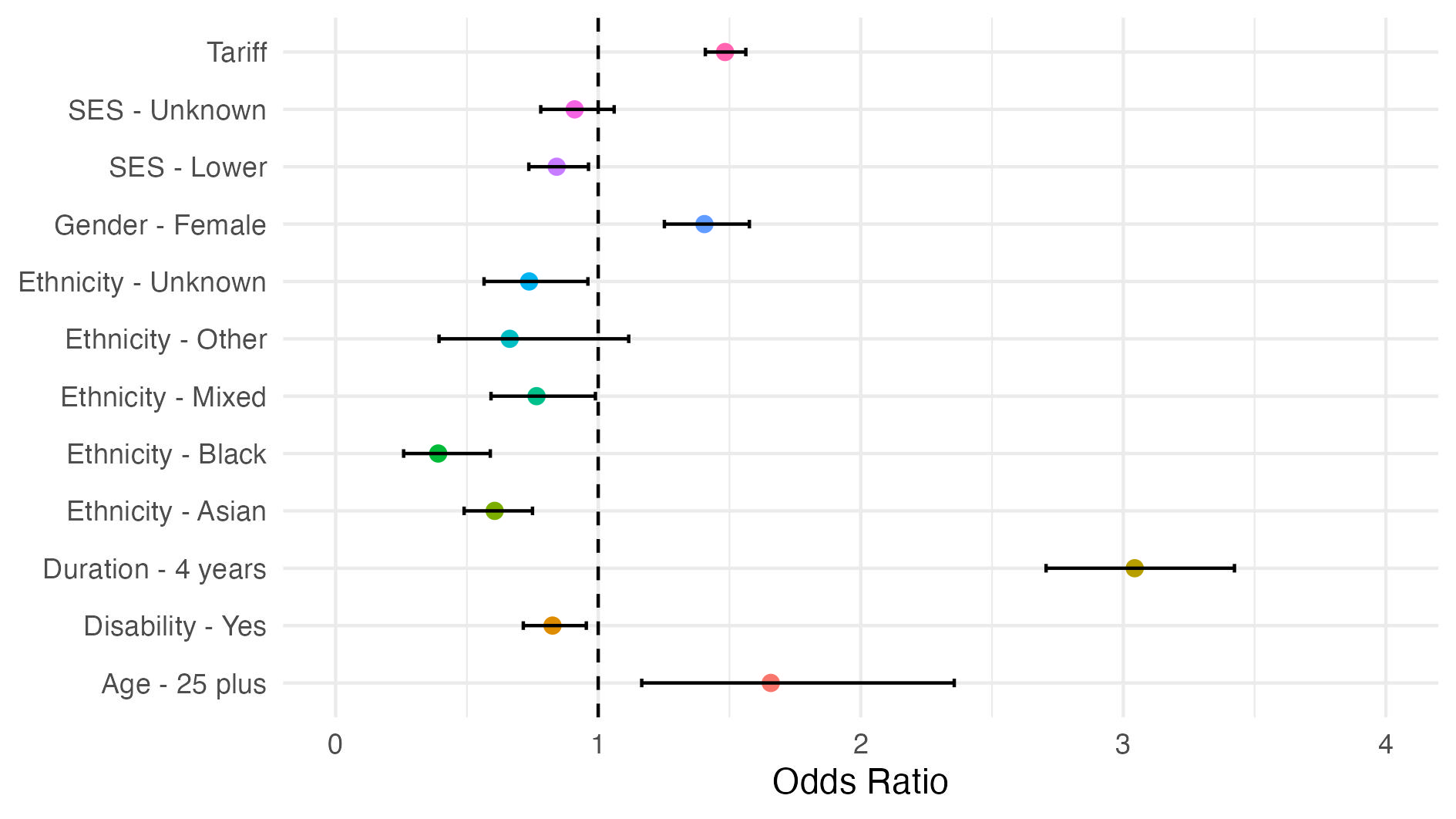}
        \caption{Forest plot showing odds ratios for coefficients of the best hierarchical model (Table \ref{tab:combined_model_summary}). Error bars represent 95\% confidence intervals. The dashed vertical reference line indicates an odds ratio of 1. Odds ratios for the \textit{Year} variable are suppressed for clarity.}
        \label{abd}
    \end{minipage}\hfill
    \begin{minipage}{0.48\textwidth}
        \centering
        \includegraphics[width=\linewidth]{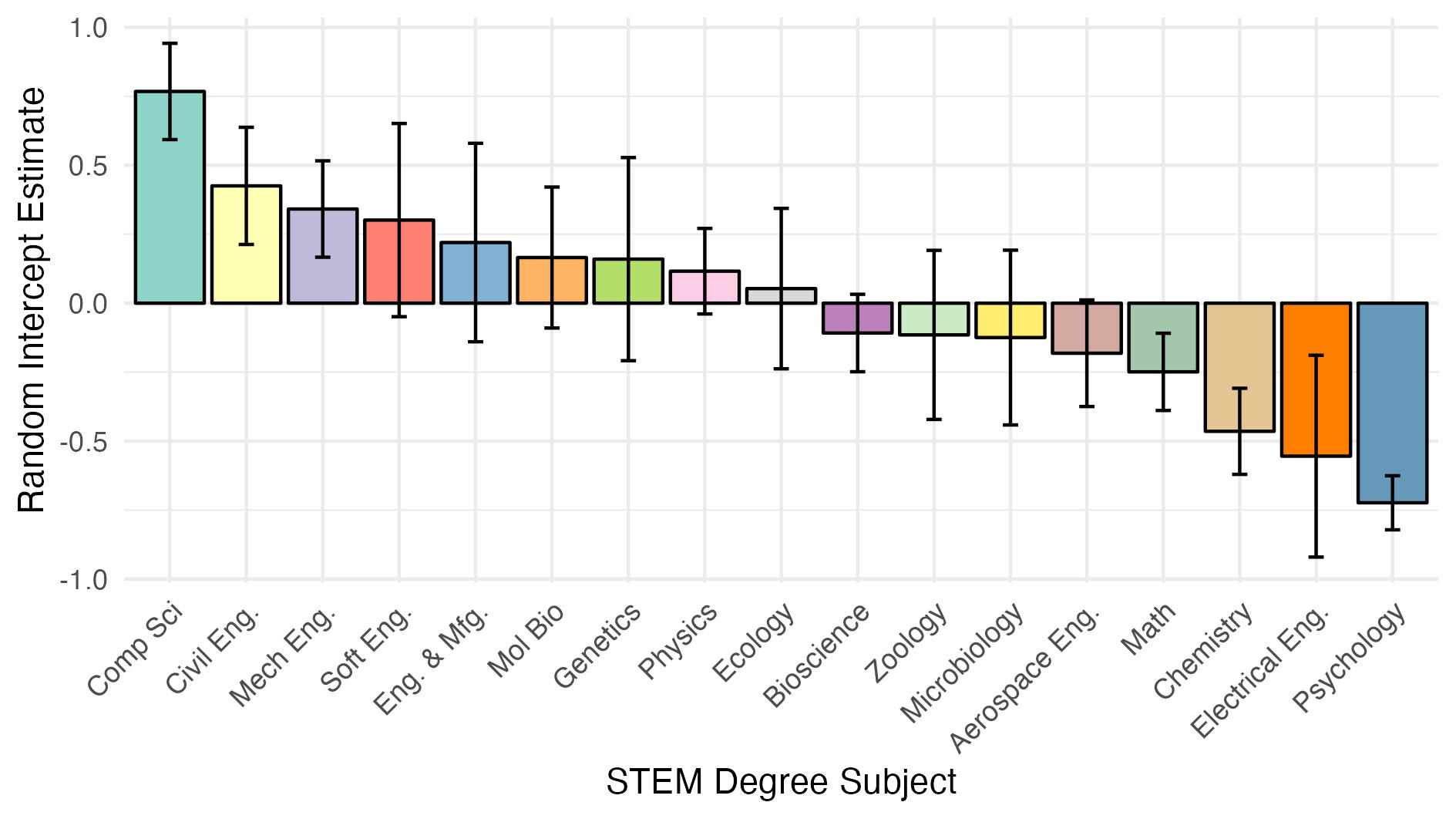}
        \caption{Plot showing the random intercept estimates for the group-level \textit{STEM Subject} variable for the best hierarchical model. The random intercept estimate for each subject represents the deviation from the fixed effects intercept estimate. Error bars represent the 95$\%$ confidence intervals of the random intercept estimate.}
        \label{randomintercept}
    \end{minipage}
\end{figure*}

\textit{\textbf{Age:}} Univariate analysis indicates that students age 25+ at graduation have statistically significantly lower mean Tariff scores compared to students age 18-24. Although the proportion of first-class degrees awarded to older students is marginally higher than that of younger students, this difference is not statistically significant. 

\textit{\textbf{Tariff:}} Interestingly, we observe that the mean UCAS Tariff score has decreased over time, with the difference compared to 2013/14 becoming statistically significantly lower from 2017/18 onwards. Simultaneously, the proportion of first-class degrees awarded has been statistically significantly higher in every year from 2015/16 onwards compared to 2013/14. These trends are illustrated in Figure \ref{fig:propovertime}. This warrants further investigation to better understand how the likelihood of achieving a first-class degree has evolved over time, particularly when controlling for UCAS Tariff scores.

These initial findings provide a critical foundation for the subsequent inferential analysis which will further explore the effects of these predictors on degree outcomes, adjusting for multiple variables and clustering of data in a multivariate hierarchical logistic regression framework.

\subsection{Predictors of First-Class Degrees and Evidence of STEM Awarding Gaps}

Here we present the key findings from our best hierarchical logistic regression model, which identifies significant predictors of achieving a first-class STEM degree, while adjusting for socio-demographic and academic factors and accounting for clustering by \textit{STEM Subject}. A summary of the full model-building process can be found in the appendix (Table \ref{HLMmodels}), and the coefficients for our best model are presented in Table \ref{tab:combined_model_summary}. Notably, \textit{Parental Ed} was excluded from the model building process because the introduction of a random intercept for \textit{STEM Subject} rendered this variable statistically insignificant, even when tested alongside other predictors. Similarly, although \textit{School} showed preliminary significance in the univariate analysis, it did not remain a statistically significant predictor in the multivariate models. Thus, both variables were omitted to achieve a more parsimonious model.

The results show clear evidence of ethnicity-based awarding gaps. After accounting for clustering by subject and controlling for UCAS Tariff score, gender, socio-economic status, disability, age, course duration, and year of graduation, Black students have 0.39 (95\% CI: 0.26–0.59) times the odds of achieving a first-class degree compared to White students. Asian students have 0.61 (95\% CI: 0.49–0.75) times the odds, while students of Mixed ethnicity have 0.77 (95\% CI: 0.59–0.99) times the odds of obtaining a first-class degree. The Average Marginal Effects further illustrate these disparities, with Black students having, on average, a 16\% lower probability (AME = -0.16, 95\% CI: -0.22, -0.10) of achieving a first-class degree compared to White students, while Asian students have an average 9\% lower probability (AME = -0.09, 95\% CI: -0.13, -0.06).

Interestingly, the gender gap reverses in the hierarchical model. While the univariate analysis (Table \ref{univariate}) shows a statistically significantly higher proportion of male students achieving first-class degrees, our best hierarchical model reveals that, after adjusting for academic and socio-demographic factors and accounting for clustering by subject, female students have 1.40 (95\% CI: 1.25–1.58) times the odds of achieving a first-class degree compared to male students. The Average Marginal Effect for female students further reflects this shift, showing that, on average, female students have a 7\% higher probability of obtaining a first-class degree than male students (AME = 0.07, 95\% CI: 0.04–0.09).

For disabled students, the disparity remains evident. Students with a declared disability have 0.83 (95\% CI: 0.71–0.95) times the odds  of achieving a first-class degree compared to those without a known disability, consistent with the univariate analysis. On average, students with a disability have a 4\% lower probability of achieving a first-class degree (AME = -0.04, 95\% CI: -0.06, -0.01).

Socio-economic status (SES) continues to be a significant factor. Students from lower SES backgrounds have 0.84 (95\% CI: 0.74–0.96) times the odds of achieving a first-class degree than their peers from higher SES backgrounds, with an average 3\% lower probability (AME = -0.03, 95\% CI: -0.06, -0.01). Students from unknown SES backgrounds show no statistically significant difference, with an odds ratio of 0.91 (95\% CI: 0.78–1.06) and an AME of -0.02 (95\% CI: -0.05, 0.01).

In contrast to the univariate analysis, age emerges as a significant predictor in the hierarchical model once Tariff is controlled for. Students aged 25 and over at graduation have 1.66 (95\% CI: 1.17–2.36) times the odds of achieving a first-class degree compared to students age 18–24. The AME for students over 25 indicates a 10\% higher probability of achieving a first-class degree compared to students age 18-24 (AME = 0.10, 95\% CI: 0.03–0.18).

Degree duration is also a strong predictor of achieving a first-class degree. Students graduating from 4-year programmes have 3.04 (95\% CI: 2.71–3.42) times the odds of attaining a first-class degree compared to those on 3-year programmes, even when controlling for UCAS Tariff score. The AME of 0.24 (95\% CI: 0.21–0.27) further highlights this difference, showing that, on average, students on 4-year programmes have a 24\% higher probability of achieving a first-class degree.

\begin{table*}[]
\centering
\caption{\label{interaction_results} Coefficient estimates for the interaction terms between each predictor and \textit{Year} on the log-odds scale. The 2013/14 row shows baseline coefficients for each predictor. For each subsequent year, coefficients represent the change in log-odds relative to 2013/14 for the interaction between \textit{Year} and the predictor listed at the top of each column. Each year’s log-odds is calculated by adding its interaction coefficient to the baseline. Standard errors are shown in parentheses. Bold values indicate statistically significant coefficients at \(p<0.05\).}

\label{Interactiontable}
\begin{ruledtabular}
\begin{tabular}{lllllllll}
\textbf{Year} & \textbf{Eth:Black} & \textbf{Eth:Asian} & \textbf{Gender:F} & \textbf{SES:Lower} & \textbf{Dis:Y} & \textbf{Age:25+} & \textbf{Duration:4} & \textbf{Tariff} \\
\hline
\noalign{\smallskip} 
13/14 & -0.04 (0.81) & -0.79 (0.55) & \textbf{0.57 (0.18)} & 0.18 (0.25) & 0.48 (0.35) & -0.65 (1.11) & \textbf{1.11 (0.19)} & \textbf{0.42 (0.09)} \\
14/15 & 0.19 (1.17) & -0.14 (0.73) & 0.18 (0.24) & -0.18 (0.35) & -0.42 (0.45) & 1.08 (1.22) & -0.30 (0.25) & 0.10 (0.12) \\
15/16 & -0.42 (1.07) & 0.92 (0.66) & 0.06 (0.24) & -0.49 (0.35) & -0.39 (0.45) & 1.89 (1.26) & -0.39 (0.25) & -0.10 (0.12) \\
16/17 & -13.19 (11.27) & 0.52 (0.63) & -0.13 (0.22) & -0.11 (0.32) & -0.67 (0.41) & 0.60 (1.25) & 0.06 (0.24) & 0.01 (0.12) \\
17/18 & -0.81 (0.97) & -0.40 (0.69) & -0.18 (0.22) & -0.31 (0.32) & -0.68 (0.41) & 0.08 (1.36) & 0.09 (0.25) & -0.05 (0.11) \\
18/19 & -1.37 (1.03) & 0.59 (0.62) & -0.32 (0.22) & -0.49 (0.32) & -0.68 (0.41) & 1.47 (1.20) & -0.11 (0.24) & -0.05 (0.11) \\
19/20 & -0.73 (0.97) & -0.06 (0.62) & \textbf{-0.57 (0.22)} & \textbf{-0.64 (0.32)} & \textbf{-0.79 (0.40)} & 0.98 (1.19) & 0.26 (0.24) & 0.02 (0.12) \\
20/21 & -0.48 (0.93) & 0.35 (0.62) & -0.37 (0.22) & -0.14 (0.31) & -0.57 (0.39) & \textbf{2.52 (1.24)} & -0.07 (0.24) & -0.13 (0.11) \\
21/22 & -1.73 (1.04) & 0.40 (0.62) & \textbf{-0.45 (0.22)} & -0.60 (0.32) & \textbf{-1.00 (0.39)} & 0.72 (1.20) & 0.33 (0.24) & 0.01 (0.12) \\
\end{tabular}
\end{ruledtabular}
\end{table*}

\begin{table*}[]
\centering
\caption{\label{ame_results} Year-specific Average Marginal Effects with standard errors in parentheses for various predictors on the likelihood of achieving a first-class degree. The AMEs shown were calculated using each individual interaction model, incorporating the interaction term coefficients from Table \ref{interaction_results} into the full model for each predictor and year. Bold values indicate statistically significant AMEs at \(p<0.05\). These results are plotted in Figure \ref{AMEYEAR}.}
\label{AMETable}
\begin{ruledtabular}
\begin{tabular}{lllllllll}
\textbf{Year} & \textbf{Eth:Black} & \textbf{Eth:Asian} & \textbf{Gender:F} & \textbf{SES:Lower} & \textbf{Dis:Y} & \textbf{Age:25+} & \textbf{Duration:4} & \textbf{Tariff} \\
\hline
\noalign{\smallskip} 
13/14 & -0.01 (0.13) & -0.11 (0.06) & \textbf{0.09 (0.03)} & 0.03 (0.04) & 0.08 (0.06) & -0.09 (0.13) & \textbf{0.21 (0.04)} & \textbf{0.07 (0.01)} \\
14/15 & 0.03 (0.15) & \textbf{-0.13 (0.05)} & \textbf{0.12 (0.03)} & 0.00 (0.04) & 0.01 (0.05) & 0.08 (0.10) & \textbf{0.15 (0.03)} & \textbf{0.08 (0.01)} \\
15/16 & -0.08 (0.11) & 0.03 (0.07) & \textbf{0.12 (0.03)} & -0.06 (0.04) & 0.02 (0.06) & \textbf{0.26 (0.13)} & \textbf{0.15 (0.04)} & \textbf{0.06 (0.01)} \\
16/17 & \textbf{-0.38 (0.03)} & -0.05 (0.06) & \textbf{0.09 (0.03)} & 0.01 (0.04) & -0.04 (0.04) & -0.01 (0.11) & \textbf{0.26 (0.04)} & \textbf{0.08 (0.01)} \\
17/18 & -0.16 (0.08) & \textbf{-0.21 (0.06)} & \textbf{0.08 (0.03)} & -0.03 (0.04) & -0.04 (0.04) & -0.10 (0.13) & \textbf{0.27 (0.04)} & \textbf{0.07 (0.01)} \\
18/19 & \textbf{-0.22 (0.07)} & -0.04 (0.05) & 0.05 (0.03) & -0.06 (0.04) & -0.04 (0.04) & 0.17 (0.10) & \textbf{0.22 (0.03)} & \textbf{0.07 (0.01)} \\
19/20 & -0.14 (0.08) & \textbf{-0.15 (0.04)} & 0.00 (0.03) & \textbf{-0.09 (0.03)} & -0.04 (0.04) & 0.07 (0.09) & \textbf{0.30 (0.03)} & \textbf{0.09 (0.01)} \\
20/21 & -0.10 (0.09) & -0.09 (0.05) & 0.04 (0.03) & 0.01 (0.04) & -0.02 (0.04) & \textbf{0.37 (0.09)} & \textbf{0.24 (0.03)} & \textbf{0.06 (0.01)} \\
21/22 & \textbf{-0.27 (0.06)} & -0.08 (0.05) & 0.02 (0.03) & \textbf{-0.08 (0.04)} & \textbf{-0.10 (0.03)} & 0.01 (0.09) & \textbf{0.32 (0.03)} & \textbf{0.09 (0.02)} \\
\end{tabular}
\end{ruledtabular}
\end{table*}

As expected, pre-university attainment in the form of UCAS Tariff score is a strong predictor of first-class degree outcomes. Each one standard deviation increase in UCAS Tariff score increases the odds of achieving a first-class degree by 1.48 (95\% CI: 1.41–1.56). The AME suggests that a one standard deviation increase in Tariff score raises the average probability of achieving a first-class degree by 8\% (AME = 0.08, 95\% CI: 0.07–0.09), after controlling for other factors.

Finally, the odds of achieving a first-class degree increased from 2015/16 onwards compared to the 2013/14 baseline year, peaking in 2020/21 during the COVID-19 pandemic with an odds ratio of 2.57 (95\% CI: 2.06–3.21). This reflects a sharp rise in the average probability of achieving a first-class degree during the pandemic (18\% higher in 2020/21 compared to 2013/14). These trends align with the univariate analysis, and the possibility of grade inflation will be discussed in more detail in the discussion section.

In terms of model fit, our best hierarchical model has a pseudo \(R^2\) value of 0.18, indicating a moderate level of explanatory power. The AUC value of 0.72 suggests that the model has good discriminative ability, effectively distinguishing between students who achieved a first-class degree and those who did not. A detailed summary of model fit metrics and diagnostics can be found in Appendix \ref{sec:model_diagnostics}.

Figure \ref{randomintercept} displays the variation in random intercepts across different STEM subjects. Each random intercept represents the deviation from the average baseline log odds of achieving a first-class degree across all subjects (when all predictor variables are set to their reference levels). For instance, Computer Science has the largest random intercept estimate (0.77, SE = 0.09, 95\% CI: 0.59 to 0.94), indicating a higher baseline likelihood of achieving a first-class degree compared to other subjects. In contrast, Psychology has the lowest baseline estimate (-0.72, SE = 0.05, 95\% CI: -0.82 to -0.63), suggesting a comparatively lower likelihood of first-class degree outcomes. The standard deviation of the random intercepts is 0.41, and the intra-class correlation coefficient (ICC) of 0.05 indicates that approximately 5\% of the total variance in first-class degree outcomes is attributable to differences between STEM subjects. This suggests that, while subject-specific effects account for some variability in first-class degree outcomes, the majority of variance is explained by individual-level factors.


\subsection{Trends in Awarding Gaps Over Time}

\begin{figure*}[tp]
        \includegraphics[width=\linewidth]{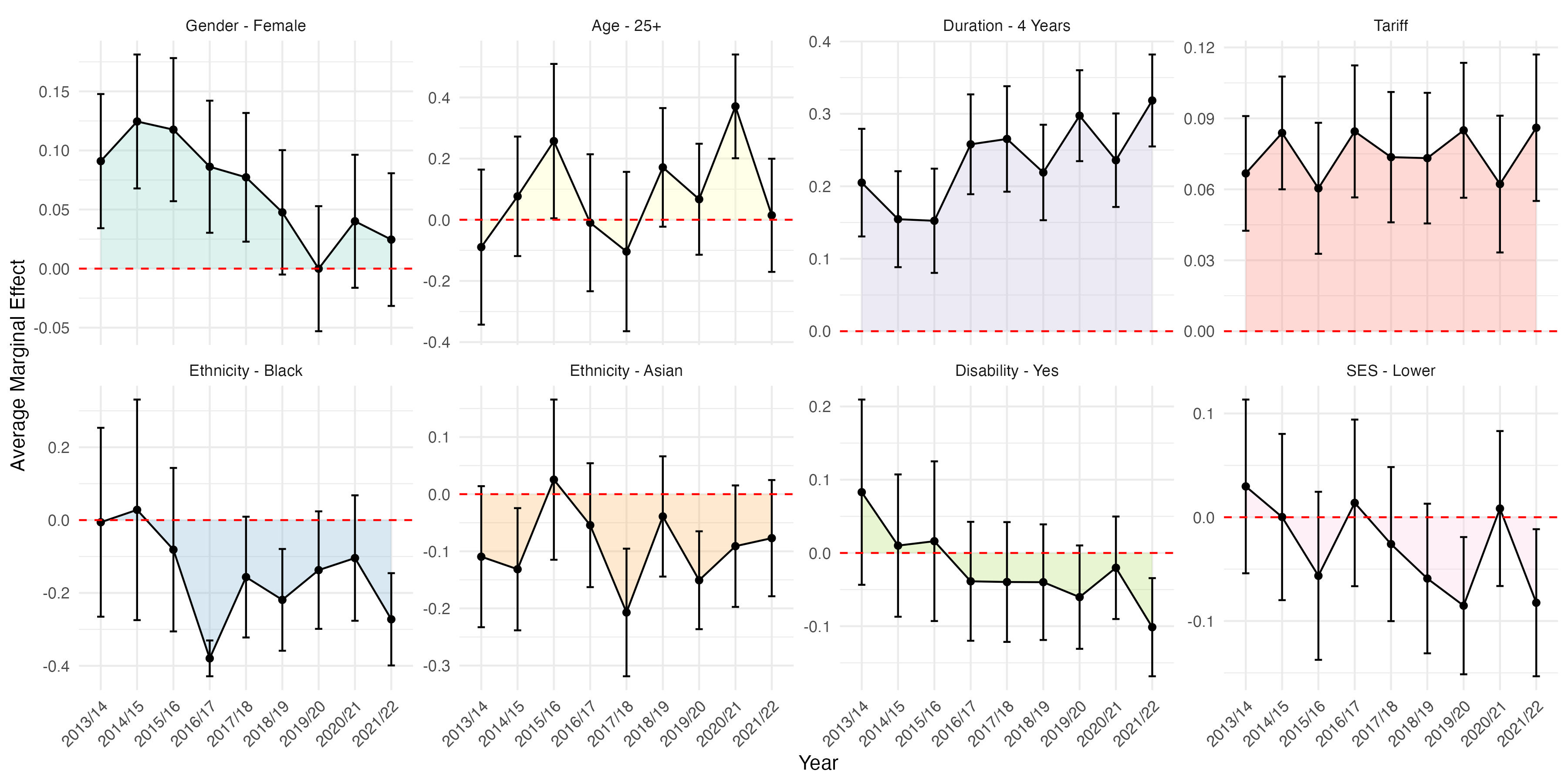}
        \caption{Average Marginal Effects (AMEs) of Socio-Demographic and Academic Predictors on the Likelihood of Achieving a First-Class Degree Across Different Years. Year-specific AMEs were estimated using the interaction models and interaction terms presented in Table \ref{Interactiontable} for each predictor (e.g., gender, age, ethnicity, pre-university tariff score, and study duration). Error bars represent 95\% confidence intervals, with a dashed line at zero indicating no effect. The shaded regions highlight the difference between a positive or negative AME and zero, which helps to visualise how awarding gaps have changed over time.}
        \label{AMEYEAR}
\end{figure*} 

To assess whether awarding gaps have changed over time relative to the baseline year (2013/14), we extended our best hierarchical model by sequentially introducing interactions between \textit{Year} and each predictor. In each iteration, we added a single interaction term (e.g. \textit{Year*Gender}) to the base model, extracted the interaction coefficients, and then removed the interaction term before testing the next variable’s interaction with \textit{Year}. This systematic approach allowed us to evaluate each predictor’s interaction with \textit{Year} independently, ensuring consistency across analyses. Table \ref{interaction_results} presents a summary of the interaction coefficients across these models. For each interaction model, we calculated year-specific average marginal effects for each predictor, summarised in Table \ref{AMETable}. The trends in awarding gaps over time are visualised in Figure \ref{AMEYEAR}, showing how each gap has evolved since 2013/14.

\textbf{Ethnicity-related Trends}: The interaction coefficients for \textit{Ethnicity-Black} and \textit{Ethnicity-Asian} with \textit{Year} indicate no statistically significant changes in awarding gaps relative to 2013/14. The AME plot provides additional insight, with large error bars for the early years (2013/14 to 2015/16) reflecting smaller sample sizes, which limits robust temporal comparisons to this baseline. From 2016/17 onward, AMEs for Black students remain consistently lower than for White students, with significant negative AMEs of \(-0.38\) (SE = 0.03) in 2016/17, \(-0.22\) (SE = 0.07) in 2018/19, and \(-0.27\) (SE = 0.06) in 2021/22. For Asian students, a generally negative trend is seen across most years, with significant negative AMEs in 2014/15 at \(-0.13\) (SE = 0.05), 2017/18 at \(-0.21\) (SE = 0.06), and 2019/20 at \(-0.15\) (SE = 0.04). Despite these yearly fluctuations, changes relative to the 2013/14 baseline are not statistically significant at the 95\% confidence level, indicating a relatively stable and persistent awarding gap for both Black and Asian students since 2013/14.

\textbf{Gender-related Trends}: The interaction between \textit{Gender} and \textit{Year} reveals notable temporal shifts in the awarding gap. Specifically, the interaction coefficients decrease after 2013/14, becoming increasingly negative from 2015/16 onward, with statistically significant interaction coefficients of \(-0.57\) (SE = 0.22, \(p < 0.05\)) in 2019/20 and \(-0.45\) (SE = 0.22, \(p < 0.05\)) in 2021/22, with a marginally significant interaction in 2020/21. The year-specific AME plot further clarifies this pattern: female students show a consistent, statistically significant positive AME for achieving a first-class degree from 2013/14 through 2017/18. For instance, the AME peaks in 2014/15 and 2015/16 at \(0.12\) (SE = 0.03, \(p < 0.05\)), indicating an average 12\% higher probability of achieving a first-class degree for female students relative to male students in these years. However, this advantage diminishes gradually after 2014/15, with AMEs statistically non-significant from 2018/19 to 2021/22. The combined evidence from both the AME plot and interaction coefficients suggests that, while female students initially had a higher probability of achieving a first-class degree, this advantage has decreased in recent years.

\textbf{Disability-related Trends}: The interaction analysis shows a variable impact of disability status on the likelihood of achieving a first-class degree over time. Interaction terms between \textit{Disability} and \textit{Year} reveal a general trend of negative values relative to the baseline year 2013/14, with statistically significant effects in 2019/20 (\(-0.79\), SE = 0.40, \(p < 0.05\)) and 2021/22 (\(-1.00\), SE = 0.39, \(p < 0.05\)). These negative coefficients indicate an increased disadvantage for disabled students in achieving first-class degrees in these years relative to 2013/14. The AMEs are consistently negative from 2016/17 onwards, aligning with the interaction terms, although only the 2021/22 AME (\(-0.10\), SE = 0.03, \(p < 0.05\)) reaches statistical significance. Figure \ref{AMEYEAR} illustrates this trend, showing that while disabled students consistently face challenges in achieving first-class degrees, these barriers were particularly pronounced in 2021/22.

\textbf{SES-related Trends}: Socio-economic status shows a complex relationship with first-class degree outcomes over time. Interaction terms for \textit{SES:Lower} and \textit{Year} are generally negative, reflecting a trend of lower odds for students from lower SES backgrounds compared to higher SES peers, though these effects are statistically significant only in specific years. In particular, the interaction coefficient in 2019/20 is \(-0.64\) (SE = 0.32, \(p < 0.05\)), suggesting that in this year students from lower SES backgrounds faced increased disadvantage in achieving a first-class degree relative to the baseline year, 2013/14. We see from Table \ref{ame_results} and the AME plot that the Average Marginal Effects for lower SES students are consistently negative across years, with statistical significance in both 2019/20 (\(-0.09\), SE = 0.03, \(p < 0.05\)) and 2021/22 (\(-0.08\), SE = 0.04, \(p < 0.05\)). This pattern suggests that socio-economic disparities in first-class degree outcomes were particularly pronounced in these years.

\textbf{Age-related Trends}: The relationship between age at graduation and first-class degree outcomes varies over time, with mostly positive AMEs indicating a slightly higher probability of achieving a first-class degree for students aged over 25. Table \ref{interaction_results} shows that interaction terms are mostly non-significant, with the exception of 2020/21, where a significant positive interaction term (\(2.52\), SE = 1.24, \(p < 0.05\)) suggests a higher likelihood of achieving a first-class degree for older students compared to 2013/14. This aligns with the AME analysis, where the AME in 2020/21 is significantly positive at \(0.37\) (SE = 0.09, \(p < 0.05\)). Another notable positive AME appears in 2015/16, with a value of \(0.26\) (SE = 0.13, \(p < 0.05\)), suggesting that these years particularly favoured older students in achieving first-class degrees.

\textbf{Duration-related Trends}: The relationship between degree duration and the likelihood of achieving a first-class degree remains stable over time, as indicated by non-significant interaction terms between \textit{Duration} and \textit{Year}. Students graduating from four-year degree programmes consistently show an advantage. Year-specific AMEs reinforce this trend, with significant values of 0.21 (SE = 0.04) in 2013/14, increasing to 0.30 (SE = 0.03) by 2019/20. These results indicate that students who complete four-year degree programmes have statistically significantly higher odds of achieving a first-class degree across all years, with this advantage remaining stable over time.

\textbf{Tariff-related Trends}: The \textit{Tariff} variable shows a strong and consistent positive relationship with first-class degree outcomes across all years in the dataset, with no statistically significant variation relative to 2013/14. Year-specific AMEs reinforce this trend, with positive values across all years, such as 0.07 (SE = 0.01, \(p < 0.05\)) in 2013/14, which increases slightly to 0.09 (SE = 0.01, \(p < 0.05\)) by 2019/20. These findings collectively indicate that higher pre-university tariff scores are robustly associated with an increased likelihood of obtaining a first-class degree, an effect that has remained stable over time.

\section{\label{sec:level1} Discussion}

We found that ethnicity, disability, gender, socioeconomic status, age, pre-university UCAS Tariff score, and year of graduation are all statistically significant predictors of achieving a first-class STEM degree at this research-intensive Russell Group university. Our analysis reveals clear awarding gaps associated with ethnicity, socioeconomic status, disability status, gender, and age, underscoring the range of demographic factors that impact first-class STEM degree outcomes. Notably, Black students face the greatest disadvantage, with 0.39 times the odds (95\% CI: 0.26–0.59) of achieving a first-class degree compared to White students, with similarly lower odds observed for other minority ethnic groups. These results align with a growing body of research documenting ethnic disparities in STEM outcomes, emphasising how systemic factors may hinder equitable access to top academic achievements for students from underrepresented ethnic backgrounds within competitive disciplines.

The challenges faced by minority ethnic students in STEM fields are well-documented, with multiple studies pointing to factors such as implicit biases, limited representation, and fewer mentorship opportunities as contributors to performance disparities. Additionally, a prevailing ``brilliance" mindset within some STEM cultures—which values innate talent over effort—may amplify these challenges by adding pressure on minority students to prove their worth, often in less supportive environments \cite{Leslie2015, Canning2019, Muradoglu2024}. Such cultural barriers not only affect students' experiences but may also impact their academic outcomes. It is therefore essential for institutions to provide targeted interventions within STEM programmes to foster a more inclusive academic culture.

\begin{figure}[tp]
    \includegraphics[width=\linewidth]{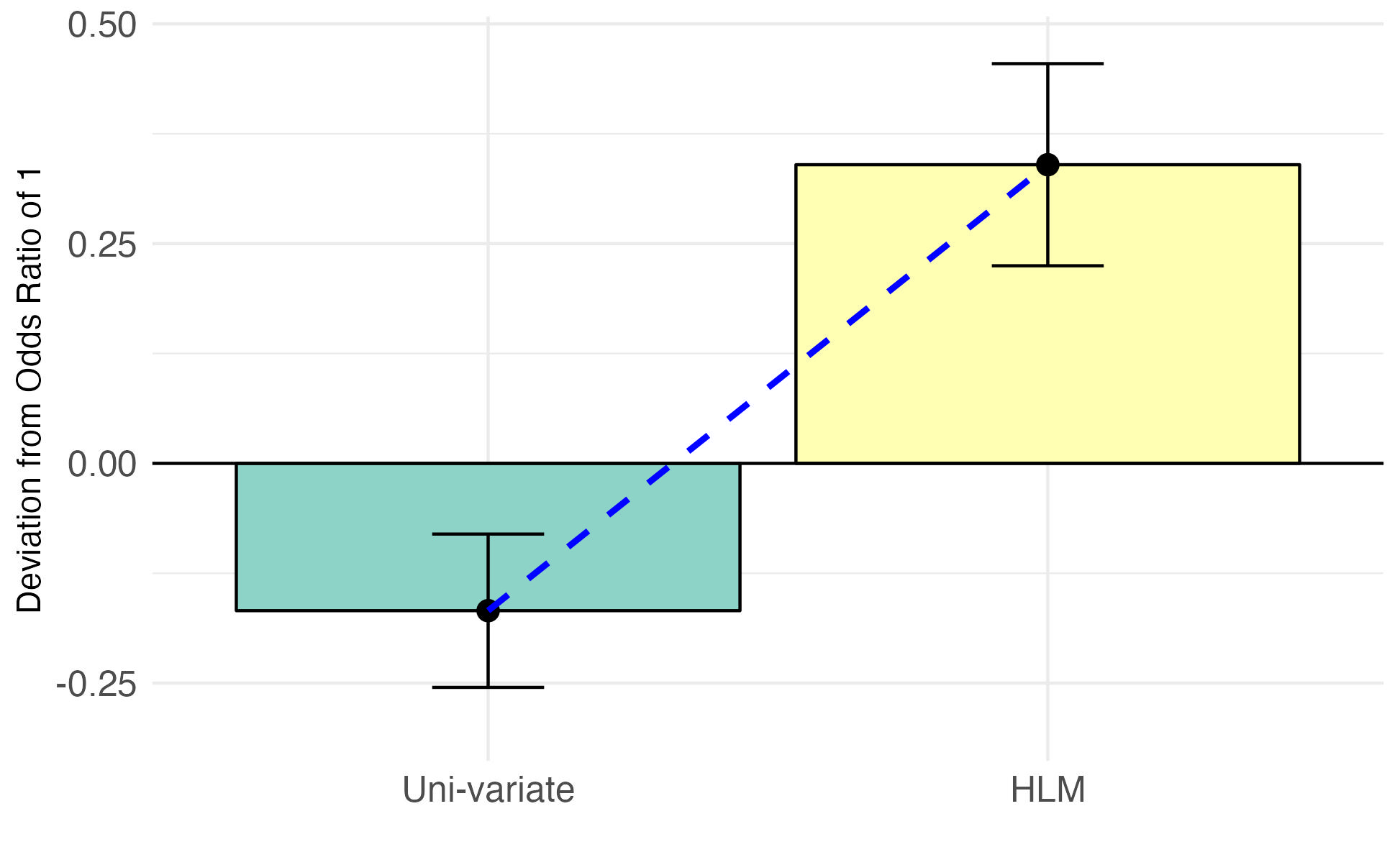}
    \caption{Comparison of odds ratios for the categorical variable \textit{Gender}. The graph shows the deviation from an odds ratio of 1 for a female student achieving a first-class degree compared to a male student, based on the univariate analysis and hierarchical linear model analysis. Error bars represent the 95\% confidence intervals for the odds ratios. The horizontal line at y = 0 represents an odds ratio of 1, indicating no difference between male and female students.}
    \label{oddscompare}
\end{figure}

Another key finding from this study is that, according to our best hierarchical model, female students have 1.40 (95\% CI: 1.25–1.58) times the odds  and an average 7\% higher probability (AME = 0.07, 95\% CI: 0.04–0.09) of achieving a first-class STEM degree compared to male students. This is despite the univariate analysis showing that the raw proportion of female students achieving a first class degree is statistically significantly lower than that of male students. Specifically, Table \ref{univariate} shows that female students have only 0.85 times the odds (95\% CI: 0.78–0.92) of achieving a first-class degree when analysed with a univariate approach. This model dependence of the gender awarding gap is also illustrated in Figure \ref{oddscompare}, where the direction and magnitude of the gap vary depending on whether a univariate or multivariate hierarchical model is applied.

This shift in reported awarding gap may stem from several key factors. Firstly, our hierarchical model includes a random intercept for each STEM subject, which accounts for differences in the baseline likelihood of achieving a first-class degree. Notably, in our dataset subjects with higher proportions of first-class degrees, such as Computer Science (49.5\%) and Physics (50.0\%), also have high male-to-female ratios (7.21 and 3.27, respectively). Conversely, fields like Psychology, which awards fewer first-class degrees (21.5\%), has a predominantly female student base (male-to-female ratio of 0.17). Additionally, female students are underrepresented in 4-year degree programmes compared to male students (male-to-female ratio of 2.16 in 4-year courses vs. 0.81 in 3-year courses), where 57.2\% of students attain first-class degrees compared to 26.5\% in 3-year programmes. After adjusting for these variables in a multivariate hierarchical model, the apparent disadvantage for female students disappears, revealing that, under comparable conditions, female students outperform male students. This finding adds nuance to the complex picture of gender-based STEM outcomes, highlighting the importance of using multivariate models and accounting for hierarchical data structures to avoid drawing misleading conclusions from raw comparisons alone. Further investigation into the unique dynamics of this institution’s STEM programmes and broader support structures may provide valuable insight into the mechanisms driving this observed gender advantage.

Our findings also revealed socioeconomic, disability, and age-related awarding gaps in achieving first-class degrees. Students from lower SES backgrounds had lower odds of achieving a first-class degree than their higher SES peers, which the literature suggests may be due to limited access to resources and academic support \cite{Smith2001, Galindo-Rueda2004}. Disability status also emerged as a significant predictor, with the multivariate hierarchical analysis revealing that students with a declared disability are significantly less likely to achieve a first-class degree than those without, underscoring potential academic and accessibility challenges impacting their performance \cite{Barnes2021}. Additionally, age played a significant role: students aged 25 and over had higher odds of achieving a first-class degree compared to younger students (aged 18–24), despite entering university with significantly lower average UCAS Tariff scores. This finding aligns with studies suggesting that older students may bring increased motivation and resilience to their studies, positively influencing academic outcomes \cite{doi:10.1080/0260293970220305, richardson1994b}.

One of the key insights from our analysis was the substantial impact of course duration on the likelihood of achieving a first-class degree. Students graduating from 4-year degrees had over three times the odds of attaining a first-class degree compared to those on 3-year degrees. While one might naively attribute this difference to higher UCAS Tariff scores required for 4-year integrated master’s courses, our findings reveal that even after controlling for Tariff scores and various socio-demographic variables, students on 4-year courses still have, on average, a 24\% higher probability of achieving a first-class degree than those on 3-year courses. This persistent advantage suggests that factors beyond academic entry standards—such as increased motivation, resilience, or the development of a professional identity over the extended programme—may contribute to the higher degree outcomes. Further investigation is needed to explore these potential mechanisms and to better understand the experiences and motivations that may drive these differences.

Our analysis also revealed a significant temporal trend, with the odds of achieving a first-class degree steadily increasing over the ten-year period, peaking during the COVID-19 pandemic in 2020/21, suggesting significant grade inflation. This trend aligns with findings from the Office for Students which indicates wide-spread unexplained grade inflation across the higher education sector over the past decade \cite{OfS2023time}. The particularly sharp spike observed during the pandemic years, especially in 2021/22, likely stems from emergency adjustments in assessment policies. Institutions widely replaced traditional exams with alternative and continuous assessment formats to mitigate the pandemic’s disruptive impact on student learning. These changes may have unintentionally led to higher grades by emphasising continuous assessment strategies, increasing flexibility in submission timelines, and potentially adjusting grading criteria to address unprecedented challenges faced by students. Additionally, our analysis hints at the fact that disparities in first-class degree outcomes based on ethnicity, disability, socio-economic status, and gender may have slightly narrowed during 2020/21. This finding is consistent with recent research indicating that assessment changes during the pandemic may have reduced the size of awarding gaps \cite{PJ2023, Cagliesi2023}.

One of the novel contributions of this study is the use of interaction terms and the calculation of year-specific AMEs to investigate how socio-demographic awarding gaps have evolved over time. The temporal patterns observed were complex: interaction terms between socio-demographic variables and year of graduation indicate that awarding gaps related to ethnicity, socio-economic status, and disability have remained relatively stable from 2013/14 to 2021/22, while the higher probability for female students (relative to males) has gradually declined. It is important to note, however, that due to relatively large uncertainties associated with the year-specific AMEs—rooted in substantial standard errors of the interaction terms, often from smaller sample sizes within individual years—it was challenging to detect statistically significant temporal shifts. In several cases, the interaction terms suggested potential changes that may have been more evident with larger samples. The general stability in socio-demographic awarding gaps over time is concerning, given the high-profile nature of these disparities and the extensive resources many universities have dedicated over the past decade to address them. Our findings indicate that, within this timeframe, implemented initiatives at this particular institution have had minimal impact on reducing these gaps. Extending this type of analysis to other institutions or conducting a multi-institutional study could reveal whether these gaps are similarly persistent elsewhere or if certain strategies have proven more effective in specific contexts. This broader perspective is essential for institutions to measure intervention impacts accurately and to share best practices.

\section{Implications}

Addressing awarding gaps requires a multifaceted approach that integrates curriculum reform, inclusive teaching practices, and support systems tailored to students’ diverse backgrounds. 

Curriculum and assessment design represent areas where universities can make impactful changes. Research suggests that traditional grading scales, such as percentage-based systems, can deepen ethnic and racial disparities in STEM courses \cite{PhysRevPhysEducRes.18.020103}. Changes to grade weighting, course structure, and content sequencing—such as covering foundational concepts before advanced applications—can reduce these disparities and promote a more inclusive academic environment \cite{PhysRevPhysEducRes.16.020125, PhysRevPhysEducRes.19.020126}. Assessment modifications, including partial credit for problem-solving and retake options, have been shown to support gender equity by mitigating test anxiety’s impact \cite{Good2022}.

Furthermore, the institutional and disciplinary culture within STEM departments, particularly the prevalence of a “brilliance” mindset that values innate talent over effort, plays a significant role in shaping student success. This cultural focus can inadvertently marginalise underrepresented groups, who may feel additional pressure to validate their abilities within unsupportive settings \cite{Muradoglu2024}. Addressing these cultural elements, including expanding mentorship, representation, and emotional support, may cultivate a more inclusive learning environment. Strategies that include financial support, diversifying curricula, and dismantling systemic barriers are also crucial to fostering equity \cite{Wong2021}.

Finally, rigorous, data-driven methods are essential for accurately measuring and monitoring awarding gaps, as they allow institutions to assess the true impact of their interventions. By using robust analytical frameworks that control for factors such as academic background, socio-demographic characteristics, and the hierarchical structure of educational data, institutions can gain clearer insights into the effectiveness of their strategies. The framework presented in this study provides a model for identifying, tracking, and evaluating these gaps over time, enabling institutions to refine their interventions based on evidence and share successful practices that promote equity.

\section{Limitations and Future Directions}

Our study is limited in several respects that warrant further discussion. Firstly, our sample consisted only of data from a single Russell Group university between 2013/14 and 2021/22, which may limit the generalisability of our findings beyond this specific institution. Secondly, the relatively small sample sizes within certain socio-demographic subgroups (e.g., minority ethnic students, disabled students, lower SES students), and within each individual year, limited our ability to obtain precise estimates for interaction terms. Large standard errors on these terms resulted in substantial uncertainty, making it challenging to identify statistically significant year-to-year changes in awarding gaps unless these changes were dramatic. This limitation underscores the importance of larger samples when analysing fine-grained temporal trends, and future research with expanded sample sizes could provide clearer insights into the dynamics of awarding gaps over time.

The self-reported nature of variables such as \textit{Disability} and \textit{Ethnicity} also introduces complexity. ‘Unknown’ categories appeared for \textit{Ethnicity}, \textit{SES}, \textit{School}, and \textit{Parental Ed}. In our univariate analysis, students with ‘unknown’ values in \textit{Ethnicity}, \textit{SES}, and \textit{Parental Ed} had significantly lower proportions of first-class degrees compared to their respective reference groups. In the full multivariate model, students categorised as \textit{Ethnicity: Unknown} continued to show lower odds of achieving a first-class degree, suggesting that students with ‘unknown’ values may experience unique or unmeasured disadvantages. Further research could clarify if ‘unknown’ designations indicate broader socio-demographic challenges.

Using Average Marginal Effects (AMEs) was valuable for providing a population-wide perspective on predictors of first-class degree outcomes, but debate remains over their suitability for hierarchical models. Alternative methods like Marginal Effects at the Mean (MEMs) or Estimated Marginal Means (EMMs) could yield further insights \cite{Heiss2022, leeper2024margins, LongFreese2006, Breen2018}. Future research comparing these methods could help verify findings across various approaches.

Additionally, exploring more complex interaction terms, such as between Ethnicity and Gender, could reveal important intersectional effects. For example, preliminary analysis using our dataset has found that female students from minority ethnic backgrounds experience greater relative disadvantages in achieving first-class degrees than their male counterparts from minority ethnic backgrounds. Similarly, preliminary tests of a model incorporating interactions between Gender and UCAS Tariff showed that prior academic preparation is a significantly stronger predictor of first-class degree outcomes for female students than for male students. These findings suggest that incorporating a wider range of interaction terms could provide a more nuanced understanding of socio-demographic influences on degree outcomes and help to refine targeted interventions.

Future research should investigate the underlying mechanisms driving disparities in STEM outcomes, with an emphasis on interventions that can be tracked over time to measure their impact. Targeting areas such as campus culture, mentoring opportunities, and inclusive curricula represents an essential next step. Additionally, incorporating qualitative research could provide valuable insights by capturing student experiences and institutional contexts that shape these disparities, complementing the quantitative findings. Expanding this analysis to a multi-institutional level would not only leverage a larger sample size but also determine whether awarding gaps persist across different settings and which interventions prove effective across institutions.

While our study focused on a binary outcome variable, it is important to recognise the limitations of this approach, along with those of other common awarding gap metrics that often oversimplify complex equity issues by relying on binary thresholds and aggregated demographic categories, potentially obscuring essential subgroup and intersectional differences \cite{Hubbard2024}. Addressing these limitations through more nuanced metrics that account for multiple outcomes—not just whether a student achieved a first-class degree—could improve the effectiveness of interventions by revealing trends that binary metrics might overlook. A natural extension of this research study would involve hierarchical ordinal regression, which captures all possible degree outcomes while accounting for confounding variables and data clustering. 

Finally, establishing consistent statistical approaches and guidelines for selecting appropriate methodologies based on sample size and data structure—for instance, determining when simple z-tests are sufficient versus when multivariate hierarchical modeling is necessary—would greatly support institutions in accurately identifying and tracking awarding gaps. This study provides a foundational framework, offering a first step toward accessible and rigorous methods that can be adapted and applied by other institutions seeking to identify and track awarding gaps.

\begin{acknowledgments}
This research was funded by a UoL Faculty of Science and Engineering Education Enhancement Fund. Data for this research were provided by Jisc, under license from the Higher Education Statistics Agency (HESA) and UCAS. Copyright Jisc 2024. Neither Jisc nor Jisc Services Limited can accept responsibility for any inferences or conclusions derived by third parties from data or other information supplied by Jisc or Jisc Services Limited. UCAS data is licensed subject to the terms of the Creative Commons Attribution 4.0 International Public License. 
\end{acknowledgments}

\appendix

\begin{table*}[tp]
\caption{\label{table4} Summary of model coefficients for six hierarchical binary logistic regression models, analysing the likelihood of achieving a first-class degree based on combinations of socio-demographic variables. UCAS Tariff score has been mean-centered. A random intercept was included for STEM degree subject. The standard error of the coefficient appears in brackets, and * \(p<0.05\), ** \(p<0.01\), and *** \(p<0.001\). Model fit metrics including the AIC, BIC, Pseudo-$R^2$, and AUC are also presented.}
\label{HLMmodels}
\begin{ruledtabular}
\begin{tabular}{lllllll}
 & Model 1 & Model 2 & Model 3 & Model 4 & Model 5 & Model 6 \\
\hline \\
\textbf{(Intercept)} & -0.55 (0.13)*** & -0.49 (0.13)*** & -0.51 (0.13)*** & -0.87 (0.11)*** & -0.88 (0.11)*** & -1.44 (0.14)*** \\
\\
\textbf{Ethnicity - Asian} & -0.60 (0.10)*** & -0.57 (0.10)*** & -0.48 (0.11)*** & -0.46 (0.11)*** & -0.45 (0.11)*** & -0.50 (0.11)*** \\
\\
\textbf{Ethnicity - Black} & -1.04 (0.20)*** & -1.02 (0.20)*** & -0.88 (0.20)*** & -0.86 (0.21)*** & -0.88 (0.21)*** & -0.94 (0.21)*** \\
\\
\textbf{Ethnicity - Mixed} & -0.31 (0.13)* & -0.30 (0.13)* & -0.23 (0.13) & -0.23 (0.13) & -0.22 (0.13) & -0.26 (0.13)* \\
\\
\textbf{Ethnicity - Other} & -0.50 (0.26) & -0.48 (0.26) & -0.43 (0.26) & -0.36 (0.26) & -0.35 (0.26) & -0.41 (0.27) \\
\\
\textbf{Ethnicity - Unknown} & -0.64 (0.11)*** & -0.55 (0.15)*** & -0.53 (0.15)*** & -0.38 (0.16)* & -0.34 (0.16)* & -0.37 (0.16)* \\
\\
\textbf{Gender - Female} & 0.38 (0.06)*** & 0.38 (0.06)*** & 0.31 (0.06)*** & 0.36 (0.06)*** & 0.36 (0.06)*** & 0.34 (0.06)*** \\
\\
\textbf{Disability - Yes} & -0.20 (0.07)** & -0.19 (0.07)** & -0.12 (0.07) & -0.10 (0.07) & -0.12 (0.07) & -0.19 (0.07)* \\
\\
\textbf{SES - Lower} & & -0.19 (0.07)** & -0.16 (0.07)* & -0.16 (0.07)* & -0.17 (0.07)* & -0.18 (0.07)* \\
\\
\textbf{SES - Unknown} & & -0.13 (0.07) & -0.15 (0.08)* & -0.15 (0.08) & -0.16 (0.08)* & -0.10 (0.08) \\
\\
\textbf{School - Private} & & -0.16 (0.08) & -0.10 (0.09) & -0.08 (0.09) & -0.08 (0.09) & -0.05 (0.09) \\
\\
\textbf{School - Unknown} & & -0.00 (0.13) & 0.12 (0.13) & 0.17 (0.13) & 0.13 (0.13) & 0.12 (0.14) \\
\\
\textbf{Tariff} & & & 0.38 (0.02)*** & 0.33 (0.03)*** & 0.33 (0.03)*** & 0.39 (0.03)*** \\
\\
\textbf{Duration - 4 years} & & & & 1.13 (0.06)*** & 1.13 (0.06)*** & 1.11 (0.06)*** \\
\\
\textbf{Age - 25 plus} & & & & & 0.43 (0.18)* & 0.48 (0.18)** \\
\\
\textbf{Year - 2014/15} & & & & & & 0.10 (0.12) \\
\\
\textbf{Year - 2015/16} & & & & & & 0.46 (0.12)*** \\
\\
\textbf{Year - 2016/17} & & & & & & 0.67 (0.11)*** \\
\\
\textbf{Year - 2017/18} & & & & & & 0.68 (0.11)*** \\
\\
\textbf{Year - 2018/19} & & & & & & 0.62 (0.11)*** \\
\\
\textbf{Year - 2019/20} & & & & & & 0.61 (0.11)*** \\
\\
\textbf{Year - 2020/21} & & & & & & 0.94 (0.11)*** \\
\\
\textbf{Year - 2021/22} & & & & & & 0.76 (0.11)*** \\ \\
\hline
\noalign{\smallskip} 
N & 8965 & 8965 & 8965 & 8965 & 8965 & 8965 \\
N (Degree Subject) & 17 & 17 & 17 & 17 & 17 & 17 \\
AIC & 11050.21 & 11045.64 & 10791.26 & 10428.85 & 10425.32 & 10326.96 \\
BIC & 11114.12 & 11137.96 & 10890.68 & 10535.37 & 10538.94 & 10497.39 \\
Pseudo $R^2$ (fixed) & 0.03 & 0.03 & 0.07 & 0.12 & 0.12 & 0.14 \\
Pseudo $R^2$ (total) & 0.09 & 0.10 & 0.13 & 0.16 & 0.16 & 0.18 \\
AUC & 0.65 & 0.66 & 0.68 & 0.71 & 0.71 & 0.72 \\
\end{tabular}
\end{ruledtabular}
\end{table*}

\section{Hierarchical Model Building}
\label{sec:hierarchical_model_building}

A summary of each hierarchical logistic regression model constructed during the model building process is provided in Table \ref{HLMmodels}. Notably, \textit{Parental Ed} is excluded from the model building process, as adding a random intercept for STEM subject rendered this variable statistically insignificant, even when tested alongside other predictors. It was therefore omitted from further model development.

 Model 1 incorporates \textit{Ethnicity}, \textit{Gender}, and \textit{Disability} as fixed effects and includes a random intercept to account for variation by \textit{STEM Subject}. In this model, all ethnicity categories, except `Other', are statistically significant: Asian, Black, Mixed, and Unknown ethnicities show lower log-odds of achieving a first-class degree compared to White students. Additionally, female students have higher log-odds than male students, while students with a declared disability have lower log-odds of achieving a first-class degree than those without a disability.

\begin{table}[bp]
\caption{\label{table_lrtest} Likelihood ratio test comparing Model 6 from Table \ref{table4} with the `best' model presented in Table \ref{tab:combined_model_summary}.}
\label{LRT}
\begin{ruledtabular}
\begin{tabular}{cccccc}
\textbf{Model} & \textbf{\#Df} & \textbf{LogLik} & \textbf{Df} & \textbf{Chisq} & \textbf{Pr($>$Chisq)} \\
\hline
\noalign{\smallskip} 
Best Model & 22 & -5140.1 & & & \\ 
Model 6 & 24 & -5139.5 & 2 & 1.2615 & 0.5322 \\
\end{tabular}
\end{ruledtabular}
\end{table}

In Model 2, categorical variables for \textit{SES} and \textit{School} type were added. Lower SES status is associated with lower log-odds of achieving a first-class degree, a statistically significant effect. However, attending a private school or having an unknown school type did not emerge as statistically significant predictors. The inclusion of these variables had minimal effect on the coefficients and significance levels of \textit{Ethnicity}, \textit{Gender}, and \textit{Disability}.

In Model 3, we included prior academic achievement using the \textit{Tariff} variable which reduced the magnitude and significance of the \textit{Disability} coefficient, rendering it non-significant. The statistical significance of \textit{Ethnicity: Mixed} also decreased, becoming non-significant. 

In Model 4, we added \textit{Duration}, which significantly improved the model fit and increased the Pseudo $R^2$. The inclusion of \textit{Duration} did not substantially alter the coefficients or significance levels of the other variables.

In Models 5 and 6, we added \textit{Age} and \textit{Year} of graduation, respectively. The addition of \textit{Age} in Model 5 showed that students aged 25 and over have higher log-odds of achieving a first-class degree, and this effect is statistically significant. In Model 6, including \textit{Year} further improved the model fit, with all years from 2015/16 onwards having statistically significantly higher log-odds of achieving a first-class degree compared to the reference year of 2013/14. Notably, in Model 6, the \textit{Disability} variable regained statistical significance, while \textit{School} remained non-significant throughout.

From Table \ref{HLMmodels}, we see that Model 6 has the lowest AIC and BIC values, as well as the highest total Pseudo $R^2$ score and AUC, indicating the best overall fit. In Model 6, \textit{School} is not a statistically significant predictor of first-class degrees,  and therefore we decided to drop \textit{School} from our final `best' model. This decision was supported by performing a likelihood ratio test (LRT) between the refined best model (Table \ref{tab:combined_model_summary}) and Model 6 from Table \ref{HLMmodels}. The LRT output in Table \ref{LRT} shows that excluding the non-significant \textit{School} variable does not significantly worsen the model fit compared to Model 6. Consequently, the more parsimonious model was chosen for this research study.

\section{Model Diagnostics}
\label{sec:model_diagnostics}

\begin{figure*}[tp]
    \centering
    \begin{minipage}{0.475\textwidth}
        \centering
        \includegraphics[width=\linewidth]{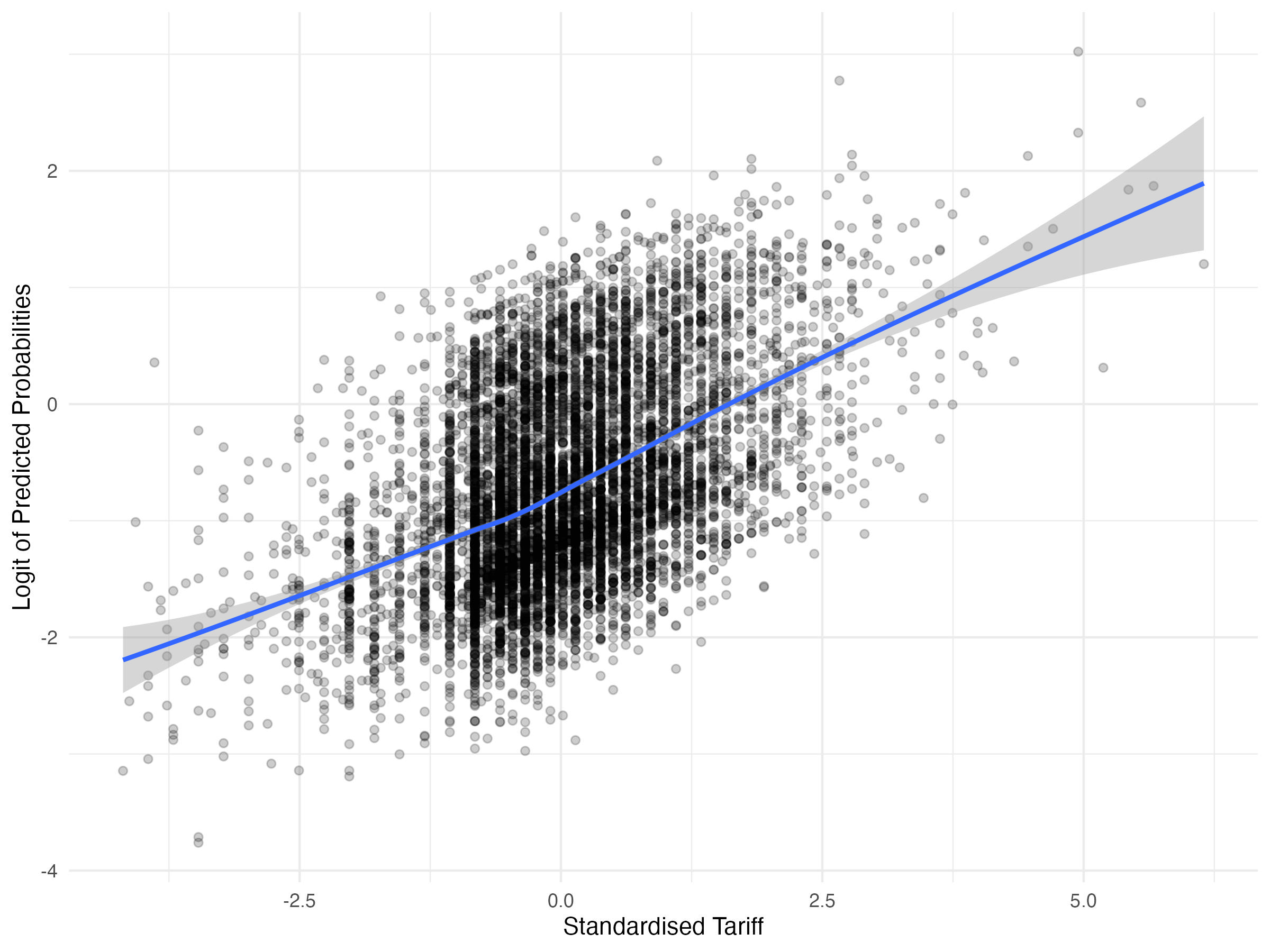}
        \caption{Plot showing the logit of the predicted probabilities resulting from the `best' model plotted against the standardised UCAS Tariff score.}
        \label{logit}
    \end{minipage}
    \hfill 
    \begin{minipage}{0.475\textwidth}
        \centering
        \includegraphics[width=\linewidth]{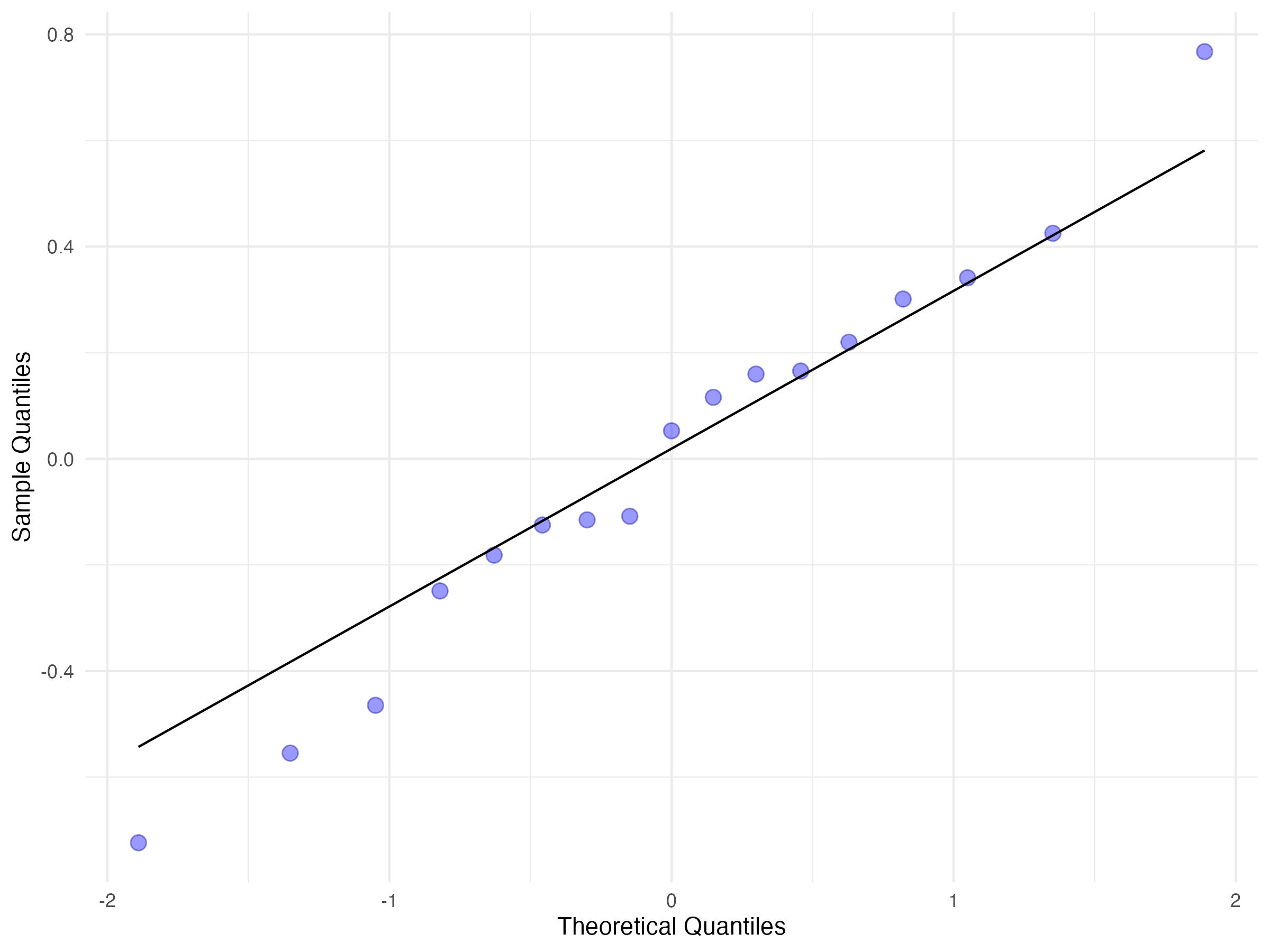}
        \caption{Q-Q plot of random intercepts corresponding to STEM degree subject.}
        \label{qq}
    \end{minipage}
\end{figure*}

In this section, we examine model diagnostics and model assumptions relating to our `best' hierarchical logistic model (Table \ref{tab:combined_model_summary}). We begin by testing for multi-collinearity using the Generalised Variance Inflation Factor (GVIF), a summary of which can be found in Table \ref{tab:gvif_summary}. We see that all adjusted GVIF values were close to 1, and well below the commonly accepted threshold of 2.5, indicating low multi-collinearity among the predictors. 

To check the linearity assumption between the logit of the predicted probabilities and the one continuous predictor variable in our model (standardised UCAS Tariff score), a plot was generated which can be seen in Figure \ref{logit}. We see that relationship between the logit of the predicted probability and the standardised Tariff score is approximately linear, suggesting that the linearity assumption is satisfied.

Random effects for the \textit{STEM Subject} variable were examined using a Q-Q plot, as shown in Figure \ref{qq}. While most points align well with the line, there are minor deviations in the lower and upper tails, where a few points fall slightly above or below the line. These deviations, however, are not severe enough to question the overall normality assumption. This conclusion is further supported by a Shapiro-Wilk test (W = 0.9839, p = 0.9843), which failed to reject the null hypothesis of normality. Together, these analyses suggest that the distribution of random intercepts for the \textit{STEM Subject} variable is well-modeled by a normal distribution, lending confidence to the validity of the model's random effects structure.

\begin{table}[tp]
\centering
\caption{Generalised Variance Inflation Factors (GVIF) and their Adjusted Values for our best hierarchical model.}
\label{tab:gvif_summary}
\begin{ruledtabular}
\begin{tabular}{lcc}
\textbf{Variable} & \textbf{GVIF} & \textbf{GVIF$^{1/(2\cdot\text{Df})}$} \\
\hline
\noalign{\smallskip} 
Ethnicity  & 1.289808 & 1.025776 \\
Disability & 1.028996 & 1.014394 \\
SES        & 1.288125 & 1.065343 \\
Gender     & 1.088929 & 1.043518 \\
Age        & 1.046662 & 1.023065 \\
Tariff     & 1.128754 & 1.062429 \\
Duration   & 1.072630 & 1.035679 \\
Year       & 1.140924 & 1.008274 \\
\end{tabular}
\end{ruledtabular}
\end{table}

The predictive performance of the model was evaluated using the Receiver Operating Characteristic (ROC) curve, yielding an Area Under the Curve (AUC) value of 0.7236. This indicates a good level of discrimination, meaning the model can adequately distinguish between students who did and did not achieve a first-class degree. Additionally, the model's accuracy was calculated as 0.7182, further supporting its predictive ability.

Residual diagnostics were performed using the DHARMa package in R \cite{hartig2022DHARMa}. The residuals versus predicted values plot showed no apparent patterns, indicating a good fit. The dispersion test yielded a dispersion statistic of 0.955 and a p-value of 0.178, indicating no significant over-dispersion. Additionally, the zero-inflation test showed a ratio of observed to simulated zeros of 1.061 and a p-value of 0.194, indicating no evidence of zero-inflation in the model. Together, these diagnostics suggest that the model is appropriately specified, with no major concerns regarding over-dispersion or zero-inflation.

The Hosmer-Lemeshow goodness-of-fit (GOF) test was conducted to assess the overall fit of the model. The test returned a chi-squared value of 6.9446 with 8 degrees of freedom and a p-value of 0.5426. Since this p-value is greater than 0.05, there is no significant evidence of a lack of fit, indicating that the model adequately fits the data.

Combining all these diagnostics and assumption tests, the hierarchical logistic regression model appears to be well-specified and performs adequately. There are no significant issues with multi-collinearity, non-linearity, over-dispersion, zero-inflation, or the overall fit of the model. The residual diagnostics support the appropriateness of the model, ensuring its stability and interpretability. We therefore proceeded to select this model for our main analysis and discussion.

\bibliography{attendance}

\end{document}